\documentclass[aps,reprint,superscriptaddress,eqsecnum,nofootinbib,amsmath,amssymb,prd,floatfix]{revtex4-2}
\usepackage[utf8]{inputenc}
\usepackage[T1]{fontenc}
\usepackage[dvipsnames]{xcolor}
\usepackage{graphicx}
\usepackage[colorlinks=true,allcolors=blue]{hyperref}
\usepackage[normalem]{ulem}

\newcommand{\UVA}{Department of Physics, University of Virginia, P.O.~Box 400714, Charlottesville, Virginia 22904-4714, USA}
\newcommand{\soton}{School of Mathematical Sciences, University of Southampton, Southampton, SO17 1BJ, United Kingdom}

\newcommand{\acc}{\mathrm{acc}}
\newcommand{\DF}{\mathrm{DF}}
\newcommand{\DM}{\mathrm{DM}}
\newcommand{\dyn}{\mathrm{dyn}}
\newcommand{\enc}{\mathrm{enc}}
\newcommand{\F}{\mathrm{f}}

\newcommand{\I}{\mathrm{i}}
\newcommand{\IN}{\mathrm{in}}
\newcommand{\ISCO}{\mathrm{ISCO}}

\newcommand{\SP}{\mathrm{sp}}
\newcommand{\sta}{\mathrm{sta}}
\newcommand{\fy}{\mathrm{2,4y}}

\begin{document}

\title{Secondary accretion of dark matter in intermediate mass-ratio inspirals:\\
\textnormal{\textit{Dark-matter dynamics and gravitational-wave phase}}}
\author{David A.\ Nichols}
\email{david.nichols@virginia.edu}
\affiliation{\UVA}
\date{\today}

\author{Benjamin A.\ Wade}
\email{baw8td@virginia.edu}
\affiliation{\UVA}

\author{Alexander M.\ Grant}
\email{a.m.grant@soton.ac.uk}
\affiliation{\UVA}
\affiliation{\soton}

\begin{abstract}
When particle dark matter is bound gravitationally around a massive black hole in sufficiently high densities, the dark matter will affect the rate of inspiral of a secondary compact object that forms a binary with the massive black hole.
In this paper, we revisit previous estimates of the impact of dark-matter accretion by black-hole secondaries on the emitted gravitational waves.
We identify a region of parameter space of binaries for which estimates of the accretion were too large (specifically, because the dark-matter distribution was assumed to be unchanging throughout the process, and the secondary black hole accreted more mass in dark matter than that enclosed within the orbit of the secondary).
To restore consistency in these scenarios, we propose and implement a method to remove dark-matter particles from the distribution function when they are accreted by the secondary.
This new feedback procedure then satisfies mass conservation, and when evolved with physically reasonable initial data, the mass accreted by the secondary no longer exceeds the mass enclosed within its orbital radius.
Comparing the simulations with accretion feedback to those without this feedback, including feedback leads to a smaller gravitational-wave dephasing from binaries in which only the effects of dynamical friction are being modeled.
Nevertheless, the dephasing can be hundreds to almost a thousand gravitational-wave cycles, an amount that should allow the effects of accretion to be inferred from gravitational-wave measurements of these systems.
\end{abstract}

\maketitle

\tableofcontents

\section{Introduction}

Astronomical measurements on galactic and larger scales have produced compelling evidence for the existence of dark matter (see, e.g.,~\cite{Bertone:2016nfn} for a review). 
The underlying particle or (quantum) field that gives rise to dark matter has not yet been identified despite a large experimental and observational research program with this aim.
This has led some to advocate in favor of searching for a wide range of possible dark matter models using a broad set of techniques to increase the chances of gaining new insight into the nature of the dark matter that pervades throughout the Universe~\cite{Bertone:2018krk}.
Following the discovery of gravitational waves by the LIGO-Virgo-KAGRA Collaboration~\cite{LIGOScientific:2021djp} (and more recently by pulsar timing arrays~\cite{NANOGrav:2023gor,Reardon:2023gzh,Antoniadis:2023rey}), 
the idea of using gravitational waves to search for the presence of  dark matter in and around compact objects became more promising~\cite{Bertone:2019irm}.
While the review~\cite{Bertone:2019irm} focuses on a broad range of dark-matter candidates (spanning 90 orders of magnitude in mass) and a variety of corresponding gravitational-wave signatures, this work will focus on cold, particle dark matter and the modifications to the gravitational-wave phase induced by the dark matter, which the Laser Interferometer Space Antenna (LISA) observatory~\cite{Baker:2019nia} could measure during its operation.

\subsection{Background and context}

The possibility of using LISA to probe the dark-matter environment of a stellar-mass compact object inspiraling into an intermediate-mass black hole (IMBH)---called an intermediate-mass-ratio inspiral (IMRI)---was explored previously by Eda and collaborators~\cite{Eda:2013gg,Eda:2014kra} (and subsequently by others, e.g.,~\cite{Yue:2017iwc,Yue:2018vtk,Edwards:2019tzf,Kavanagh:2020cfn,Coogan:2021uqv,Becker:2021ivq,Becker:2022wlo,Speeney:2022ryg,Cole:2022fir,Cole:2022ucw}).
For dark matter to have a significant effect on the orbital dynamics, Eda \textit{et al}.\ found that the dark matter needed to form an overdensity,\footnote{The fact an overdensity was required was also noted in~\cite{Barausse:2014tra} in the context of environmental effects in EMRIs.} which they called a dark-matter ``minispike'' (or ``spike'' for short).
Working within the context of Newtonian physics, Eda \textit{et al}.\ identified two physical effects, in fact, that could cause the dark-matter spike to change the orbital dynamics of the IMRI strongly enough that it would have an observable effect on the emitted gravitational waves.
The first~\cite{Eda:2013gg} was a (quasi\nobreakdash-)conservative effect: the dark matter enclosed within the orbit of the secondary changes the Keplerian frequency at a given orbital radius from that of a black hole in vacuum.
While this would be challenging, at a fixed orbital radius, to distinguish from an IMRI with a slightly more massive primary, the enclosed mass changes as the system inspirals.
This (slow) time dependence of the enclosed mass leads to a small, but distinctive, change in the evolution of the frequency from what would be expected for an IMBH without surrounding dark matter.
This, in turn, produces a dephasing of the inspiral with respect to an inspiral in vacuum, which could be measured by LISA~\cite{Eda:2013gg}.
However, it was later noted by Eda \textit{et al}.\ in~\cite{Eda:2014kra} (see also~\cite{Macedo:2013qea} in a different context), that this enclosed-mass effect is, in fact, smaller than another effect that arises from the presence of dark matter known as dynamical friction~\cite{Chandrasekhar1943a}.
Dynamical friction is produced by the gravitational scattering of dark-matter particles with the secondary, which induces an overdensity (a wake) that gives rise to an effective drag force (which then increases the rate of inspiral of the IMRI system). 
Under the assumption that the distribution of dark matter remained unchanged throughout the IMRI's inspiral, some features of the density of dark matter could be inferred precisely from a gravitational-wave measurement of the system by the LISA detector~\cite{Eda:2014kra}.

The assumption that the dark-matter distribution remained static throughout the inspiral was shown in~\cite{Kavanagh:2020cfn} to be in tension with energy conservation for a nontrivial portion of the parameter space of IMRIs and dark-matter spikes studied in~\cite{Eda:2014kra}.\footnote{More specifically, the gravitational scattering that gives rise to dynamical friction is a conservative process for the combined system of the binary and dark matter.
This implies that the decrease in orbital energy of the secondary, as it inspirals because of dynamical friction, must be balanced by an increase in the energy of the dark-matter particles. 
Ref.~\cite{Kavanagh:2020cfn} showed that the energy increase was sufficiently large to unbind the entire dark-matter spike with significant kinetic energy for a large region of the IMRI (specifically, mass ratio) and dark-matter (specifically, density normalization and radial power law) parameter space.}
As a result, it was demonstrated in~\cite{Kavanagh:2020cfn} that it is necessary to jointly evolve the IMRI's orbital dynamics with the distribution of dark matter to determine consistently the effect of the dark matter on the emitted gravitational waves.
We will refer to this evolution of the dark-matter distribution in response to dynamical friction as ``\textit{dynamical-friction (DF) feedback}'' in the rest of this paper.
A procedure was developed in~\cite{Kavanagh:2020cfn} to evolve the dark matter on timescales that are long compared to the IMRI's orbital timescale, under the assumption that the dark-matter halo remained spherically symmetric by rapidly equilibrating on the orbital timescale.
The results of jointly evolving the IMRI and the dark matter were pronounced: For example, for certain representative IMRIs and dark-matter spikes considered in~\cite{Eda:2014kra}, the number of gravitational-wave cycles of dephasing (from a similar IMRI in vacuum) for a dynamically evolved dark-matter distribution could be as much as 100 times smaller than the equivalent dephasing for a distribution that remained static~\cite{Kavanagh:2020cfn}.
Nevertheless, it was shown in~\cite{Coogan:2021uqv} that despite the much smaller dephasing, the presence of dark matter around the IMRI was detectable with LISA (and waveform models with dark matter had significantly higher Bayes factors over the best-fit waveform models without dark matter); in addition, the properties of the initial dark-matter distribution could still be inferred from the gravitational waves even for events near the threshold of detection.

Several years after the work of Eda \textit{et al}., Yue and Han~\cite{Yue:2017iwc} noted that if the secondary were a black hole, there would be one additional (relativistic) effect that the dark matter would have on the rate of inspiral of the binary: the black hole would accrete dark matter as the dark-matter particles fell through the event horizon, and the secondary black hole would increase in mass.\footnote{Refs.~\cite{Eda:2013gg,Eda:2014kra,Kavanagh:2020cfn,Coogan:2021uqv} assumed that the secondary was a neutron star (and that the cross section between dark matter and nuclear matter in neutron stars was sufficiently small) so that dark-matter accretion by the neutron star would be negligible.}
To avoid confusion with accretion of dark matter onto the primary (which we will not model in this paper), we will sometimes refer to this accretion onto the secondary during the inspiral as \textit{secondary accretion (SA)} henceforth.
Ref.~\cite{Yue:2017iwc} considered the dark-matter distribution to be static as the secondary inspirals and found that it induced a dephasing with respect to vacuum systems that was a few to a few tens of a percent of the amount of the dephasing induced by dynamical friction.\footnote{As will be discussed in more detail in Sec.~\ref{sec:DMIMRI}, given the form of the accretion term in the IMRI's equations of motion, it will produce an effect on the IMRI's dynamics at one post-Newtonian (PN) order (an additional power of $v^2/c^2$) higher than the term responsible for dynamical friction. Given that the PN parameter is of order $10^{-2}$ to $10^{-1}$ at the initial frequencies for the systems considered, the size of this effect is then consistent with the order-of-magnitude expectations from a simple counting of PN orders.
While the counting of PN powers is useful for understanding the relative importance of different terms in the equations of motion, the phenomenon of accretion of dark matter by the secondary is a genuinely relativistic effect that arises because the secondary has a horizon; it is not a weak-field phenomenon.}
There is then an interesting numerical coincidence that 
the dephasing induced by dynamical friction with feedback onto the distribution function turns out to be comparable to the dephasing induced by accretion for a static halo for some binaries.
It is then natural to wonder if the effects of accretion would be comparable if they were computed in a dark-matter distribution with dynamical-friction feedback rather than a static distribution.  

We give a brief argument here why we expect the accretion in a halo with DF feedback not to change the mass accreted significantly. 
First, it is important to note that the decreased dephasing from dynamical friction, when including DF feedback, occurred because the local density of dark-matter particles moving more slowly than the orbital speed of the secondary (those particles that contribute to dynamical friction) was significantly decreased (often by one or more orders of magnitude)~\cite{Kavanagh:2020cfn,Coogan:2021uqv}.
These more slowly moving particles compose roughly half the total density in these systems, for static dark-matter distributions governed by a single power-law in radius~\cite{Kavanagh:2020cfn}.\footnote{With the more complicated functional form of the distribution function with feedback, the total local density can deplete by more than a factor of two; we will return to this point later in this paper.}
While the accretion cross section does have a dependence on velocity, the secondary black hole can accrete particles with any speeds that come close enough to the secondary's event horizon.
As a result, this suggests that the calculations of secondary accretion in~\cite{Yue:2017iwc} for static halos could be within a factor of a few of the capture that occurs in the dynamical halos of~\cite{Kavanagh:2020cfn}.

This, however, will pose a problem for computing reasonable estimates of dark-matter accretion onto the secondary during the inspiral for the following reason:
Because dark-matter accretion occurs most efficiently for dark-matter particles that are closer to the secondary's event horizon, then as the secondary inspirals, the total amount of dark matter accreted should not be much larger than the total amount of dark matter enclosed within the initial orbit.
However, as we will show in this paper, the model of dark-matter accretion used in~\cite{Yue:2017iwc} predicts for static dark-matter spikes and for less-extreme IMRI mass ratios that the mass accreted by the secondary can exceed the mass enclosed within the orbit.
From the arguments above, incorporating DF feedback will reduce the amount of accretion by only a factor of a few. 
Thus, it will be necessary develop a procedure that accounts for the loss of dark-matter particles from the distribution function as they are accreted by the secondary to ensure that mass is conserved and to obtain accurate estimates of the amount of gravitational-wave dephasing induced by accretion.
We introduce such a procedure in this paper, and it will follow in the same spirit of the DF feedback of~\cite{Kavanagh:2020cfn} (in particular, in terms of its assumption of spherical symmetry and fast equilibration over the orbital timescale).
We will call this new feedback ``\textit{secondary-accretion (SA) feedback}'' to distinguish it from the dynamical-friction feedback of~\cite{Kavanagh:2020cfn}.

Secondary accretion and dynamical friction also will ultimately have different effects on the dark-matter distribution after the secondary has inspiraled through the dark matter distribution and merged with the primary.
In Refs.~\cite{Kavanagh:2020cfn,Coogan:2021uqv}, while dynamical friction did unbind a small fraction of the dark matter particles, it primarily redistributed them to higher-energy bound orbits.
Thus, while there could be a large transient redistribution of the dark-matter particles during the inspiral, the effect on the distribution of dark matter afterwards was not particularly large.
However, because secondary accretion simply removes particles from the distribution function, it has the potential to produce a larger, lasting change in the dark matter density during and after the inspiral.
This could impact the evolution of other IMRI systems that might form subsequently.

\subsection{Summary of results and structure of this paper}

The structure of this paper, and the main results in each of the paper's parts, will now be summarized below.
Section~\ref{sec:DMIMRI} is primarily a review, in which we introduce some of the notation and approximations that we use throughout the paper in Sec.~\ref{subsec:Notation}.
We then give the evolution equations for the IMRI's orbital dynamics and the equation that determines the mass accreted onto the secondary in Sec.~\ref{subsec:IMRIeqs}.
Next, Sec.~\ref{sec:staticAccretion} gives analytical expressions for the accreted mass normalized by the enclosed mass for binaries that evolve in a static halo under the influence of either gravitational radiation reaction or both radiation reaction and dynamical friction.
We show that, in both cases, the amount of mass accreted can exceed the mass enclosed; this gives the first indication that the dark-matter distribution should be evolved self-consistently with the accretion onto the secondary.

The next part of the paper, Sec.~\ref{sec:DFfeedback}, focuses on reviewing the DF feedback procedure of~\cite{Kavanagh:2020cfn} (in Sec.~\ref{subsec:DFfeedbackReview}), and then computing new results with and without accretion onto the secondary.
The results without accretion (in Sec.~\ref{subsec:DFnoSA}) evolve the same equations of motion for the dark matter and IMRI as in~\cite{Kavanagh:2020cfn}, but they simulate larger secondary masses that had not been studied in works with DF feedback (though they had been studied for static dark-matter distributions in~\cite{Yue:2017iwc}).
While the results are qualitatively similar to those with a lighter secondary at a fixed mass ratio, they are quantitatively different, and they also cover a different range of less extreme mass ratios than those in~\cite{Kavanagh:2020cfn}.
They will then also serve as an important set of baseline simulations against which we compare the number of gravitational-wave cycles when accretion effects are included.
The final part of this section (Sec.~\ref{subsec:DFplusSAonly}) then treats secondary accretion with DF feedback but without SA feedback.
For the less-extreme mass ratios, the amount of mass accreted can be comparable to the mass enclosed within the initial orbital radius, even though it did not exceed this value for the mass ratios that we simulated.
Nevertheless, the fact that they are of the same order gives the second indication that feedback onto the dark-matter distribution in response to the mass accreted will be necessary in several cases to avoid overestimating the impact of accretion on the gravitational waves emitted from these systems (and more generally, to conserve mass during the evolution). 

The secondary-accretion-feedback formalism is introduced in Sec.~\ref{sec:accretionFeedback}, and it is applied to compute results with SA feedback in isolation.
The modifications to the evolution equations for the dark-matter distribution are derived in Sec.~\ref{subsec:SAformalism}, where they are also shown to lead to a mass loss rate from the dark-matter distribution that is equal and opposite to the rate of mass accreted by the secondary.
The effect of SA feedback on the distribution function is computed analytically in Sec.~\ref{subsec:SAonly} when the binary evolves under radiation reaction and dynamical friction is neglected.
This allows the dark-matter density to be computed efficiently and studied in more detail.
For example, it allows us to determine reasonable initial conditions corresponding to an inspiral from a larger initial radius and to determine how large the effects of accretion could be on the dark-matter distribution after the merger.

The next part (Sec.~\ref{sec:AllFeedback}) finally presents the results of simulations that include both dynamical-friction and secondary-accretion feedback.
The first part (Sec.~\ref{subsec:AllGWmacc}) discusses the gravitational-wave dephasing and the accreted mass normalized by the enclosed mass.
The dephasing induced by dynamical-friction and secondary-accretion feedback have effective post-Newtonian (PN) orders that differ from the respective static cases because of the relevant local densities for accretion and dynamical-friction effects differ when dark-matter dynamics is included.
The normalized accreted mass, with DF and SA feedback, is now smaller: at most one quarter in the cases we considered.
In Sec.~\ref{subsec:AllDMdensity}, we show the effect of SA feedback on the dark-matter distribution when combined with DF feedback.
We find that SA feedback can deplete the dark-matter density significantly even with DF feedback, though the amount of depletion is not quite as strong as that with only SA feedback in Sec.~\ref{subsec:SAonly}.
We provide further discussion and our conclusions in Sec.~\ref{sec:conclusions}.

\section{Dark matter in intermediate-mass-ratio inspirals and the binary's evolution} \label{sec:DMIMRI}

This section will begin by reviewing some notation that we use to describe the orbital dynamics of the IMRI and the dark-matter distribution.
We then turn to the equations of motion that describe the IMRI's dynamics (including the effects of dark-matter accretion, as in~\cite{Yue:2017iwc}).

\subsection{Notation and approximations used} \label{subsec:Notation}

As in~\cite{Kavanagh:2020cfn}, we will denote the dark-matter density of a spherically symmetric, power-law profile by 
\begin{equation} \label{eq:static_DM}
    \rho_\DM(r) = \left\{ \begin{array}{ll} 
    \rho_\SP (r_\SP/r)^{\gamma_\SP} & r_\IN \leq r \leq r_\SP \\
    0 & r < r_\IN
    \end{array} \right. .
\end{equation}
This density is what was referred to as a ``dark-matter spike'' in~\cite{Eda:2014kra}.
For a dark-matter distribution formed during the adiabatic growth of a smaller seed black hole into an IMBH, the power law index $\gamma_\SP$ is in the range $(9/4,9/2)$~\cite{Gondolo:1999ef}.
As in~\cite{Kavanagh:2020cfn}, however, we often will allow for a wider range of possible values for $\gamma_\SP$ in case the formation scenario does not precisely match this adiabatic-growth prescription (which can be disrupted by a number of processes~\cite{Ullio:2001fb}).
The inner radius is assumed to be $r_\IN = 4 G m_1/c^2$, where $m_1$ is the mass of the IMBH (the primary), which is the inner radius at which the density goes to zero in the relativistic calculations of dark-matter spikes in~\cite{Sadeghian:2013laa}.\footnote{We do not incorporate any of the other relativistic features found in the density in~\cite{Sadeghian:2013laa} in the region closest to the black-hole horizon in our Newtonian analysis.
We use the prescription of a sharp cut at $r_\IN$, as was done in Refs.~\cite{Kavanagh:2020cfn,Coogan:2021uqv}, so as to more easily compare with the results given there.
The density computed in~\cite{Sadeghian:2013laa} does not have this feature; it smoothly goes to zero as the radius approaches $r_\IN$.}
The distance $r_\SP$ was assumed in~\cite{Eda:2014kra} to be given by $r_\SP \approx 0.2 r_\mathrm{h}$, where $r_\mathrm{h}$ is the radius at which the enclosed dark-matter mass is twice the primary's mass:
\begin{equation} \label{eq:r_h}
    \int_{r_\IN}^{r_\mathrm{h}} \rho_\DM(r) d^3 x = 2 m_1 \, .
\end{equation}
Given $r_\SP \approx 0.2 r_\mathrm{h}$, Eq.~\eqref{eq:r_h} and the form of $\rho_\DM(r)$ in Eq.~\eqref{eq:static_DM}, this implies that $m_1$, $\rho_\SP$, $r_\SP$ and $\gamma_\SP$ are not all independent.
We will then determine $r_\SP$ from the other three variables as
\begin{equation}
    r_\SP \approx \left[ \frac{0.2^{3-\gamma_\SP}(3-\gamma_\SP) m_1}{2\pi \rho_\SP} \right]^{1/3} \, .
\end{equation}
For $r > r_\SP$, the spike would smoothly transition to the initial dark-matter halo out of which it was adiabatically compressed. 
We will not treat the binary dynamics or the dark matter at radii of $r > r_\SP$, however, which is why we do not specify the functional form of the dark-matter distribution when $r>r_\SP$ in Eq.~\eqref{eq:static_DM}.

We will also find it useful to have an expression for the mass enclosed within a given radius $r$ for a power-law density.
We write the result as in~\cite{Kavanagh:2020cfn} as
\begin{equation} \label{eq:mDM_enc}
    m_\enc(r) = \left\{ \begin{array}{ll} 
    m_\DM(r) - m_\DM(r_\IN) & r_\IN \leq r \leq r_\SP \\
    0 & r < r_\IN
    \end{array} \right.
\end{equation}
with
\begin{equation}
    m_\DM(r) = \frac{4\pi \rho_\SP r_\SP^{\gamma_\SP}}{3-\gamma_\SP} r^{3-\gamma_\SP} \, .
\end{equation}

Next, we will discuss our notation for the various masses that we will use and some common approximate expressions for these quantities.
In addition to the mass of the IMBH, $m_1$, which we have already introduced, we will denote the mass of the secondary as $m_2$.
The mass ratio will be denoted by $q = m_2 / m_1$, the total mass by $M = m_1 + m_2$, the reduced mass by $\mu = m_1 m_2 /M$, the symmetric mass ratio by $\eta = \mu/M$, and the chirp mass by $\mathcal M = \eta^{3/5} M$.
The IMRIs in this paper have $q \ll 1$.
We will work to leading order in $q$ throughout this paper, so we will frequently make approximations such as
\begin{subequations}
\begin{align}
    M = {} & m_1 (1 + q) \approx m_1 \, , \\
    \mu = {} & \frac{m_2}{1 + q} \approx m_2 \, , \\
    \eta = {} & \frac{q}{(1+q)^2} \approx q \, , \\
    \mathcal M  = {} & \frac{m_1 q^{3/5}}{(1+q)^{1/5}} \approx m_1 q^{3/5} \, .
\end{align}
\end{subequations}

It will also be useful to consider an effective (radius-dependent) mass ratio which is the ratio of the enclosed mass Eq.~\eqref{eq:mDM_enc} to the primary mass, $m_1$: 
\begin{equation}
    q_\enc(r) = m_\enc(r)/m_1 \, .
\end{equation}
We will work with dark-matter spikes, binary mass ratios $q$, and orbital separations $r_2$, for which $q_\enc(r_2) / q < 1$.
Thus, in addition to working to the leading order in $q$, we consider effects related to $q_\enc(r_2)$ to occur at an equivalent order to those in $q$ (i.e., we effectively will treat $q$ and $q_\enc(r_2)$ as the same small parameter).

Because $m_2 \approx \mu$ will be a function of time when accretion occurs, so too will $q$, $\eta$, and $\mathcal M$.
However, since the amount of mass accreted will be a small fraction of $m_2$ for the cases we consider, when we refer to $q$ (for example) throughout this paper, we will typically be referring to the initial value, which we will sometimes denote as $q_\I$ to make this explicit.
If we were to consider accretion onto the primary, $m_1$ would also be time dependent; however, we will not consider such accretion in this paper.\footnote{Ignoring accretion onto the primary can be argued to be reasonable for the following reasons.
In the absence of the secondary, accretion of dark matter onto the primary is expected to be negligible given the weak interactions between dark-matter particles.
In the presence of the secondary, some dark matter will be scattered onto orbits that can be captured by the primary black hole.
Since this requires reasonably strong scattering, the mass of the dark-matter particles accreted onto the primary during the inspiral should be of the order of $m_\enc(r_{2,\I})$, so that the fractional change in mass $m_1$ during the inspiral is of order $q_\enc(r_{2,\I})$, for $r_{2,\I}$ being the initial separation.
However, since we will not treat effects of order $q_\enc$ in this paper, neglecting accretion onto the primary should be consistent with this approximation used throughout this paper.}

When we treat time-dependent dark-matter densities, we will denote them by $\rho_\DM(r,t)$.
In the dynamical case, we will continue to assume that the density will be zero within the inner radius, and we will not evolve them at radii $r > r_\SP$ either (and we will use the same values of $r_\IN$ and $r_\SP$ as in the corresponding static density, which is used as initial data for the evolution of the time-dependent density).
We will similarly denote the mass enclosed by $m_\enc(r,t)$ in the time-dependent case.
On occasion, we also use the notation $m_\enc^\sta(r)$ for the static case and $m_\enc^\dyn(r)$ for the time-dependent case to distinguish the two without specifying the time dependence of the function in the dynamic case.
We could similarly make a time-dependent definition of $q_\enc(r,t)$, but we do not use it in this paper.

\subsection{IMRI evolution equations} \label{subsec:IMRIeqs}

The orbital dynamics of the IMRI are naturally expressed in terms of the motion of the reduced mass $\mu \approx m_2$ in the Newtonian limit.
In vacuum, the dynamics of these systems [and the closely related extreme mass-ratio inspirals (EMRIs)] are most precisely modeled using the techniques associated with the gravitational self-force (see, e.g.,~\cite{Barack:2018yvs} for a review).
When dark matter is included, however, relativistic analyses of such binaries have been more limited (see~\cite{Speeney:2022ryg} for a notable exception for static dark-matter distributions), and the description of the binary and dark matter has largely been restricted to the mostly\footnote{The ``mostly'' caveat here is that the effects of gravitational radiation reaction were taken into account using the leading (Newtonian) quadrupole formula.
This is a relativistic effect, but it is being treated at leading order using Newtonian information about the binary's quadrupole moment.} Newtonian approximation (e.g.,~\cite{Eda:2013gg,Eda:2014kra,Yue:2017iwc,Yue:2018vtk,Edwards:2019tzf}).
This is especially true of when the dark matter has been allowed to evolve in response to dynamical-friction feedback (e.g.,~\cite{Kavanagh:2020cfn,Coogan:2021uqv,Becker:2021ivq,Becker:2022wlo,Cole:2022ucw,Cole:2022fir}), as there have been no such relativistic studies (to the best of our knowledge).
We will thus not attempt to move beyond this mostly Newtonian approximation in our calculations in this paper, as we will also be considering dark-matter dynamics and feedback (both dynamical-friction and secondary-accretion feedback).

As in~\cite{Kavanagh:2020cfn}, only IMRIs in circular orbits will be treated in this paper. 
While gravitational radiation~\cite{Peters:1963ux,Peters:1964zz} and dynamical friction~\cite{Becker:2021ivq} have circularizing effects in IMRIs, the precise formation scenarios of these binaries could lead to residual eccentricity (so this assumption should be revisited in future work).
In the absence of dissipative effects (which includes dynamical friction and dark-matter accretion, in this discussion), the orbital dynamics for circular orbits is simple, as it is determined by just the Keplerian frequency
\begin{equation} \label{eq:Omega}
    \Omega = \sqrt{\frac{G[M+m_\enc(r_2)]}{(a_2)^3}} \approx \sqrt{\frac{Gm_1}{(r_2)^3}} \, ,
\end{equation}
where $a_2$ is the semi-major axis of the binary's reduced mass, and $r_2$ is the coordinate distance of the secondary from the center of mass.
The speed of the secondary in these circular orbits is given by the expression
\begin{equation} \label{eq:v_of_r2}
    v_2 = \sqrt{\frac{GM}{r_2}} \approx \sqrt{\frac{Gm_1}{r_2}} \, .
\end{equation} 
Because $r_2$ is constant on the orbital timescale, then $\Omega$ is constant as well, and the angular component of Newton's second law becomes trivial for circular orbits.

Including dark-matter accretion, dynamical friction, and gravitational radiation reaction causes the orbits to evolve on timescales much longer than the orbital timescale, so we will introduce a (slow) time dependence to $r_2(t)$ and $\Omega(t)$.
This leads to dissipative dynamics which describe how the system moves from larger to smaller circular radii.
These dynamics were computed in~\cite{Kavanagh:2020cfn} by postulating that energy balance holds. 
In~\cite{Kavanagh:2020cfn}, the change in the orbital energy of the IMRI was equated to the energy radiated from the system in gravitational waves plus the energy transferred to the dark-matter distribution through dynamical friction.
However, this approach does not work as straightforwardly when including dark-matter accretion, because, roughly speaking, the accretion can be considered an inelastic scattering process, which conserves momentum but not necessarily energy.

Here we will work directly with Newton's equations and allow the change in momentum of the secondary to have a term that arises from the increase in inertia of the secondary as it accretes dark matter (as in~\cite{Yue:2017iwc}).
Radiation reaction, dynamical friction, and accretion introduce effective forces into the tangential component of Newton's second law (in the orbital plane of the binary) that cause $\Omega$ to evolve on a timescale much longer than the orbital one.
Using the expression for this component of the acceleration, $r\dot\Omega + 2 \dot r \Omega$, and Eq.~\eqref{eq:Omega}, the tangential acceleration term reduces to $\Omega \dot r/2$, for circular orbits.
This allows us to compute an equation for the evolution of $\dot r$.
Including gravitational radiation reaction, dynamical friction, and dark-matter accretion leads to an equation of motion of the form
\begin{subequations} \label{eq:dot_r}
\begin{equation}
    \dot r_2 = -\dot r_2^\mathrm{RR} -\dot r_2^\DF - \dot r_2^\mathrm{A}
\end{equation}
where
\begin{align} \label{eq:r_dot_RR}
    \dot r_2^\mathrm{RR} & \approx q \frac{64 }{5c^5} \left( \frac{G m_1}{r_2} \right)^3 \, ,\\
    \label{eq:r_dot_DF}
    \dot r_2^\DF & \approx q \, 8\pi \sqrt{\frac{G}{m_1}} r_2^{5/2} \log\Lambda \rho_\DM(r_2, t; v < v_2)   \, ,\\
    \label{eq:r_dot_A}
    \dot r_2^\mathrm{A} & \approx 2 r_2 \dot m_2/m_2 \, .
\end{align}
\end{subequations}
The expression for $\dot r_2^\mathrm{A}$ can also be understood as the change in the orbit that occurs when the mass of the secondary increases while the orbital angular momentum remains an adiabatic invariant~\cite{Hughes:2018qxz}.
In the expression for $\dot r_2^\DF$, we introduced the notation $\rho_\DM(r_2, t; v < v_2)$ to denote the (in general, time-dependent) density of dark-matter particles at $r_2$ moving more slowly than the orbital speed of the secondary, $v_2$, and $\log\Lambda$ for the Coulomb logarithm.
As in~\cite{Kavanagh:2020cfn}, we will assume $\Lambda = \sqrt{1/q}$, where $q$ here is the initial mass ratio of the binary.\footnote{Recall that when there is accretion onto the secondary, the mass ratio will be time dependent.
However, because the mass accreted during the inspiral will be at most a few percent for the binaries we consider in this paper, we do not expect that keeping $\Lambda$ constant produces any significant errors here. 
When we numerically solve for the orbital dynamics in subsequent sections of this paper, we will not assume that $\Lambda$ is constant.}
For static dark-matter spikes of the form in Eq.~\eqref{eq:static_DM}, the fraction of dark-matter particles moving more slowly than the orbital speed at each radius is proportional to the total dark-matter density at that radius: namely $\rho_\DM(r_2; v < v_2) = \xi \rho_\DM(r_2)$ (see~\cite{Kavanagh:2020cfn}) for all radii $r_2$.\footnote{Eddington inversion relates the spherically symmetric density in Eq.~\eqref{eq:static_DM} to the distribution function in~Eq.~\eqref{eq:static_f}. From integrating Eq.~\eqref{eq:static_f} over velocities up to the orbital speed of circular orbits at $r_2$, it follows that $\xi$ is a constant that depends on just $\gamma_\SP$ which is given by
\begin{equation}
    \xi = 1 - I_{1/2}(\gamma_\SP-1/2,3/2) \, .
\end{equation}
The function $I_{1/2}(\gamma_\SP-1/2,3/2)$ is the regularized incomplete beta function.
For $\gamma_\SP = 7/3$, this gives rise to the value $\xi \approx 0.58$ used in Ref.~\cite{Kavanagh:2020cfn}.}
In this special case, then, there is a single, constant $\xi$ that determines the fraction of particles moving more slowly than the local orbital speed at $r_2$, though for more general densities and distribution functions, this will not be the case.
To solve Eq.~\eqref{eq:dot_r}, we need to determine an evolution equation for $\dot m_2$.

For this evolution equation, we use a similar treatment of the accretion of dark matter and the evolution of the mass that was used in~\cite{Yue:2017iwc}.
Specifically we compute $\dot m_2$ from
\begin{equation} \label{eq:m2_dot}
    \dot m_2 = \sigma(v_2) \rho_\DM(r_2, t) v_2 \, ,
\end{equation}
where $\sigma(v_2)$ is the accretion cross section of a (non-rotating) black hole, and $v_2$ (the speed of secondary) is used as a proxy for the relative speed of the particles with respect to the black hole (as in~\cite{Yue:2017iwc}).
Because Ref.~\cite{Yue:2017iwc} assumed a static dark-matter density, $\rho_\DM$ was a function of only $r_2$, whereas we allow it to be a function of time, instead.
It is worth noting that, unlike with dynamical friction, the full density $\rho_\DM(r_2, t)$ contributes, regardless of the speed of the dark-matter particles.
When we specialize to static dark-matter distributions for our analytical calculations, we will find it convenient to use Eq.~\eqref{eq:m2_dot} with $\rho_\DM(r_2, t) \rightarrow \zeta \rho_\DM(r_2)$, where $\zeta$ will be a phenomenological parameter that represents the effect of being able to accrete only a fraction of the density $\rho_\DM(r_2)$.
The cross-section $\sigma(v_2)$ was computed in full general relativity in~\cite{Unruh:1976fm}; here we will only use the leading ``Newtonian'' expression 
\begin{equation} \label{eq:sigma_of_v}
    \sigma(v_2) = \frac{16\pi (G m_2)^2}{(c v_2)^2} \, . 
\end{equation}

Combining Eqs.~\eqref{eq:v_of_r2}, \eqref{eq:m2_dot}, and \eqref{eq:sigma_of_v} gives an evolution equation for $m_2$ in terms of $r_2$:
\begin{equation} \label{eq:dm2dt}
    \dot m_2 \approx 16\pi (Gm_2)^2 \rho_\DM(r_2, t) \frac{1}{c^2} \sqrt{\frac{r_2}{Gm_1}} \, .
\end{equation}
Thus, we can write the rate of change of radius due to dark-matter accretion as
\begin{equation} \label{eq:r2dotC}
    \dot r_2^\mathrm{A} \approx q \frac{32\pi}{c^2} (G r_2)^{3/2} \sqrt{m_1} \rho_\mathrm{DM}(r_2, t) \, .
\end{equation}
For our analytical calculations with static dark-matter distributions, we replace $\rho_\DM(r_2, t)$ with $\zeta \rho_\DM(r_2)$ in Eq.~\eqref{eq:r2dotC}.
The expression for $\dot r_2^\mathrm{A}$, like $\dot r_2^\mathrm{RR}$ and $\dot r_2^\DF$, naturally has a factor of $q$ that can be scaled out, which implies that it is also a small effect in the small mass-ratio expansion.
Finally, note that the ratio of $\dot r_2^\mathrm{A}$ to $\dot r_2^\DF$ is given by
\begin{equation}
    \frac{\dot r_2^\mathrm{A}}{\dot r_2^\DF} = \frac{4 G m_1}{c^2 r_2} \frac{\rho_\DM(r_2, t)}{\rho_\DM(r_2, t; v<v_2)\log\Lambda} \, .
\end{equation}
The factor of $G m_1/(c^2 r_2)$ [or equivalently $(v_2/c)^2$] shows that accretion causes a change in the orbital separation at one post-Newtonian order higher than that caused by dynamical friction.
For static dark-matter distributions, the coefficient multiplying the PN parameter is $4\zeta/(\xi\log\Lambda)$, which will typically be an order one quantity.
However, for time-dependent dark-matter densities, the ratio of the total density $\rho_\DM(r_2, t)$ to the density of particles moving more slowly than the local orbital speed $\rho_\DM(r_2, t; v<v_2)$ could be large, in which case the post-Newtonian suppression of accretion could be outweighed by the greater density available to accretion.
This scenario will arise in our discussion in Secs.~\ref{subsec:DFplusSAonly} and~\ref{subsec:AllGWmacc}.

We conclude this part with a comment about the relative perturbative orders of the different dissipative effects (radiation-reaction, dynamical friction, and dark-matter accretion) in terms of the PN parameter $(v_2/c)^2 \sim Gm_1/r_2$, the mass ratio $q$, and the enclosed mass ratio $q_\enc(r_2)$.
All effects enter at order $q$ in the mass ratio relative to the conservative dynamics associated with the Keplerian motion. 
From a simple counting of powers of $r_2$, dynamical friction (respectively, accretion) would be a \textit{negative} $11/2-\gamma_\SP$ (respectively, $9/2-\gamma_\SP$)-order effect relative to radiation reaction in the dissipative dynamics of the binary (as noted in~\cite{Kavanagh:2020cfn,Coogan:2021uqv}).
Because dynamical friction is a Newtonian effect, and radiation reaction is a 2.5PN-order effect, then it is somewhat more natural to think of it as a relative, negative 2.5PN-order effect in the radial motion for circular binaries (because of its Newtonian nature).
While having, in this sense, a negative PN order, dynamical friction can be comparable in magnitude to radiation reaction at a given orbital separation, because dynamical friction contains an additional factor of $q_\enc(r_2)$, which scales as $r_2^{3-\gamma_\SP}$ (and which accounts for the remaining negative $3-\gamma_\SP$ orders in the PN-parameter counting in terms of $r_2$).
Thus, it is more natural to think of it as a negative 2.5PN-order effect relative to radiation reaction that is one order higher in $q_\enc(r_2)$. 
A similar line of logic would also lead to thinking about dark-matter accretion as a negative 1.5PN-order effect relative to radiation reaction that is one order higher in $q_\enc(r_2)$. 
This is consistent with dark-matter accretion being 1 PN order higher than dynamical friction in the PN expansion.

\section{Enclosed dark-matter mass and its accretion for static dark-matter distributions} \label{sec:staticAccretion}

We introduce and discuss several different analytical estimates and calculations of the captured mass in this section for static dark-matter distributions (as in~\cite{Yue:2017iwc}).
We show that there are binaries for which more dark matter would be accreted onto the secondary during the inspiral than there is dark matter enclosed within the secondary's initial orbit, when assuming that the dark-matter spike remains static throughout the inspiral.

When the dark-matter distribution remains static during the inspiral, the accreted mass onto the secondary can be computed analytically in terms of elementary or special functions, respectively, in the cases in which the evolution of $r_2$ is driven either (i) by gravitational radiation reaction alone, or (ii) by both radiation reaction and dynamical friction.
In both cases, the equation for the evolution of $m_2$ in Eq.~\eqref{eq:dm2dt} can be integrated by using the chain rule 
\begin{equation} \label{eq:m_2_r_2}
    \dot m_2 = \dot r_2 \frac{dm_2}{dr_2} \, ;
\end{equation} 
this allows us to combine Eqs.~\eqref{eq:dot_r} and~\eqref{eq:dm2dt} to obtain a separable ordinary differential equation for $dm_2/dr_2$ in terms of $r_2$.
It will be useful to write the result of solving the differential equation for $dm_2/dr_2$ in terms of the ratio of $m_{2,\F} \equiv m_2(r_{2,\F})$ to  $m_{2,\I} \equiv m_2(r_{2,\I})$.
The quantity 
\begin{equation}
    m_\acc \equiv \Delta m_2 =  m_{2,\F} - m_{2,\I} ,
\end{equation}
which is the accreted mass, will also prove useful for interpreting the expressions that we derive.

For the analytical estimates given in this section, we will use the parameter $\zeta$ defined in Sec.~\ref{sec:DMIMRI} when it takes on two different values (primarily for illustrative purposes). 
We will consider both when the density is the total density at $r_2$ ($\zeta=1$) and when it is the density of particles moving more quickly than the local orbital speed for a static spike ($\zeta=1-\xi$).
The $\zeta=1$ case acts as an upper limit of the amount of dark-matter mass accreted.
The $\zeta=1-\xi$ case is an analytical estimate (for static halos) of the effect of dynamical-friction feedback taking away all of the available more slowly moving particles on the accretion process.
As we will show later in Sec.~\ref{sec:DFfeedback}, it turns out to be a good lower limit when DF feedback does not induce a large (transient) change in the distribution function; however, when there is significant dynamical-friction feedback, it is not a lower limit.

\subsection{Gravitational radiation reaction only}

Supposing that the term $\dot r^\mathrm{RR}_2$ is the only term on the right-hand side in Eq.~\eqref{eq:dot_r}, then by integrating Eq.~\eqref{eq:m_2_r_2}, we find the ratio  $m_{2,\F}/m_{2,\I}$ is given by
\begin{equation} \label{eq:m2f}
    \frac{m_{2,\F}}{m_{2,\I}} = \exp\left[-\frac{5\pi}{4 m_1} \left(\frac{c^2}{Gm_1}\right)^{3/2} \! \! \! \frac{ \zeta \rho_\SP r_\SP^{\gamma_\SP} }{9/2-\gamma_\SP} \Delta (r_{2}^{9/2-\gamma_\SP}) \right] .
\end{equation}
Here $\Delta (r_{2}^{9/2-\gamma_\SP}) \equiv r_{2,\F}^{9/2-\gamma_\SP} - r_{2,\I}^{9/2-\gamma_\SP}$ is negative for $r_{2,\F} < r_{2,\I}$, so that $m_{2,\F} > m_{2,\I}$.

To understand some features of Eq.~\eqref{eq:m2f}, we will take the limit in which the argument of the exponential is small, so that $e^{x} \approx 1+x$ is a good approximation.
Working also under the assumption that $r_{2,\I} \gg r_{2,\F} \approx r_\IN$ as well, we can write the ratio of the accreted mass to the enclosed dark-matter mass\footnote{We will normalize by the total enclosed mass in both the $\zeta=1$ and $\zeta=1-\xi$ cases, although one could argue it would be more reasonable to normalize by the enclosed mass of particles moving more quickly than the orbital speed in the $\zeta=1-\xi$ case.} as
\begin{equation} \label{eq:m_cap_by_m_enc_approx}
    \frac{m_\acc}{m_\enc(r_{2,\I})} \approx \frac 5{16} q_\I \zeta\left(\frac{3-\gamma_\SP}{9/2-\gamma_\SP}\right) \left( \frac{Gm_1}{c^2 r_{2,\I}} \right)^{-3/2} \, ,
\end{equation}
where $q_\I = m_{2,\I}/m_1$ refers to the initial mass ratio.
Again when $r_{2,\I} \gg r_{2,\F} \approx r_\IN$, it can be shown that at a fixed time from merger, $r_{2,\I}$ scales with $m_1$ and $q_\I$ as $q_\I^{1/4} m_1^{3/4}$ (assuming that leading, Newtonian-order radiation reaction is driving the inspiral).
Thus, because the mass ratio $q_\I$ scales as $1/m_1$ for a given secondary mass $m_2$, then $m_\acc/m_\enc(r_{2,\I})$ scales as $m_1^{-7/4}$.
Furthermore, it has a weak dependence on $\rho_\SP$ and $\gamma_\SP$, even though both $m_\enc(r_{2,\I})$ and $m_\acc$ depend strongly on $\rho_\SP$ and $\gamma_\SP$, as can be seen from the approximate expression in Eq.~\eqref{eq:m_cap_by_m_enc_approx}.

\begin{figure}[t!]
    \centering
    \includegraphics[width=\columnwidth]{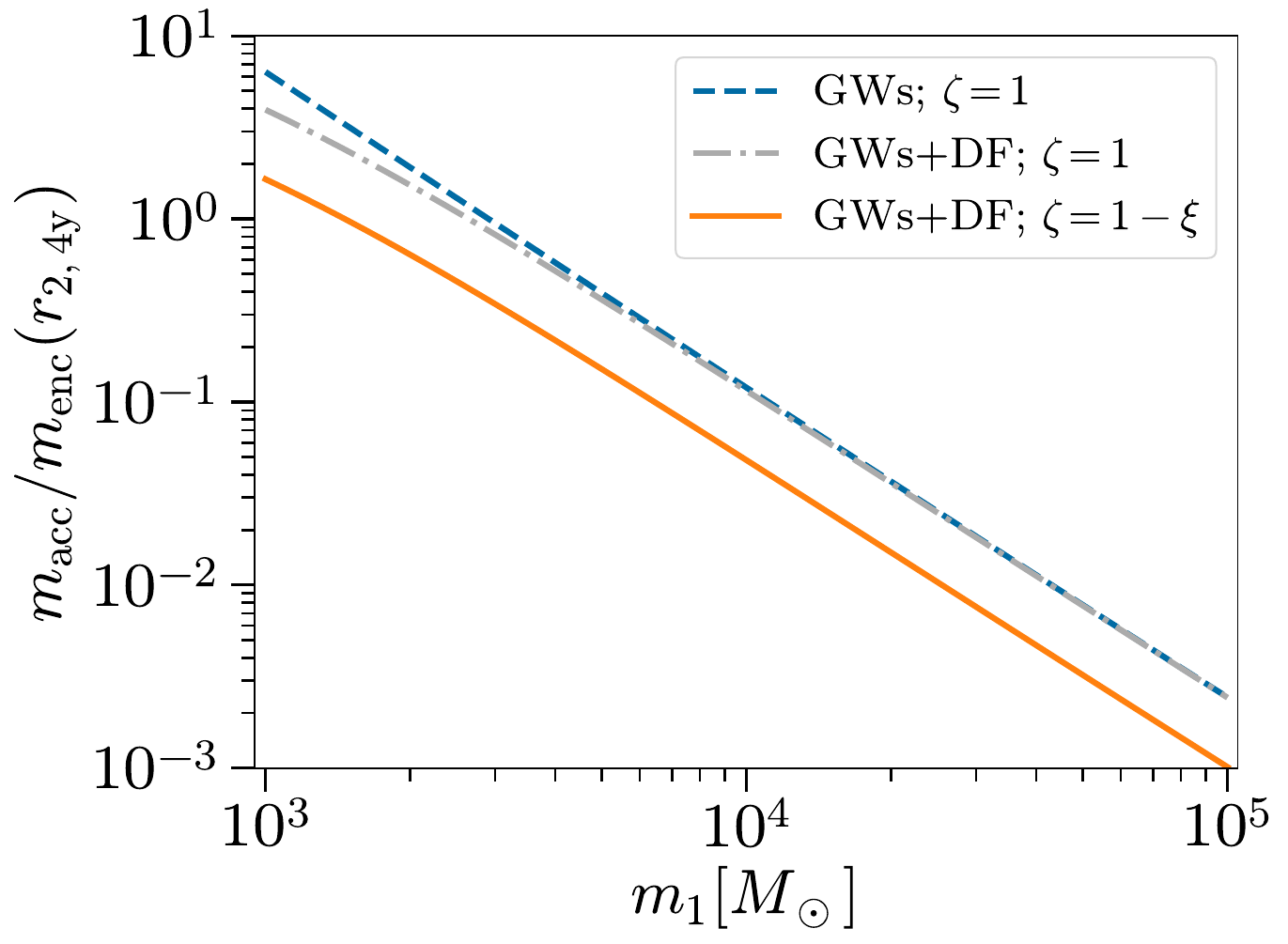}
    \caption{\textbf{The accreted mass $\mathbf{m_{acc}}$ normalized by the enclosed mass $\mathbf{m_{enc}(r_{2,4y})}$}.
    The blue dashed curve corresponds to including only gravitational radiation reaction with $\zeta=1$ corresponding to no additional restriction on the speeds to dark-matter particles being accreted.
    The gray dashed-dotted curve with $\zeta=1$ is the corresponding one with both radiation reaction and dynamical friction.
    Finally, the solid orange curve also includes dynamical friction and radiation reaction in the dynamics, but assumes $\zeta=1-\xi$, which mimics the effects of only being able to accrete particles in a static spike that move more quickly than the orbital speed of the secondary.
    The secondary mass was chosen to be $m_2=10 M_\odot$.
    The dark matter distribution is given by the static spike in Eq.~\eqref{eq:static_DM} with $\rho_\SP = 226 M_\odot/\mathrm{pc}^3$ and $\gamma_\SP = 7/3$. 
    The enclosed mass is computed for an initial orbital separation $r_\fy$ such that the IMRI will merge in four years (a duration which is consistent with the nominal length of the LISA mission).
    For all cases, there is a range of parameter space in which the accreted mass exceeds the enclosed mass.
    Further discussion of the implications of this figure is given in the text of Sec.~\ref{sec:staticAccretion}.}
    \label{fig:m_cap_by_m_enc}
\end{figure}

Because of this argument, we plot only the dependence of $m_\acc/m_\enc(r_{2,\I})$ on $m_1$ in Fig.~\ref{fig:m_cap_by_m_enc} at fixed $m_2=10 M_\odot$ for a given value of $\rho_\SP = 226 M_\odot/\mathrm{pc}^3$ and $\gamma_\SP = 7/3$.
We choose $r_{2,\I} = r_\fy$, where we use $r_\fy$ to denote the binary separation such that the secondary will inspiral to the innermost stable circular orbit (ISCO) in 4 years for each value of $m_1$.
The period of four years is chosen to be consistent with the nominal LISA mission lifetime.
Choosing different values for the parameters $\rho_\SP$ and $\gamma_\SP$ only makes a very small difference for the case of an inspiral driven by radiation reaction.
We do use the full expression for $m_{2,\F}$ in Eq.~\eqref{eq:m2f} when computing $m_\acc/m_\enc(r_\fy)$, rather than the approximate expression in~\eqref{eq:m_cap_by_m_enc_approx}.

All three curves in Fig.~\ref{fig:m_cap_by_m_enc} suggest that the previous estimates of the effect of dark-matter accretion on the IMRI's dynamics (and thus the emitted gravitational waves) were likely overestimated for smaller primary masses $m_1$.
We focus here on the dashed blue curve in which only the effects of radiation reaction are included in the evolution of $r_2$ (the other curves will be discussed in Sec.~\ref{subsec:staticGWplusDF} next).
In cases in which $m_\acc/m_\enc(r_\fy) > 1$, then more dark matter would be accreted during the inspiral than there is dark matter enclosed within a sphere of the size of the secondary's orbital radius.
For primary masses less than a few times $10^3 M_\odot$, there is likely more dark-matter accretion than there is nearby dark matter to accrete.
Moreover, even for $m_\acc/m_\enc(r_\fy) < 1$ (but not $\ll 1$), the secondary accretion could significantly alter the dark-matter distribution in the neighborhood of the secondary.

However, this estimate of accreted dark matter for a static halo when the binary evolves due only to radiation reaction (the dashed blue curve in Fig.~\ref{fig:m_cap_by_m_enc}) is an upper bound on the amount of accretion that could occur.\footnote{The reason it is an upper bound is that neglecting dynamical friction causes the binary to inspiral more slowly, thereby giving the secondary more time to accrete dark matter.
This will be shown more quantitatively in the next part, Sec.~\ref{subsec:staticGWplusDF}.}
As we show next, including the effects of dynamical friction on the binary's orbit for a static halo also can lead to a somewhat smaller estimate.

\subsection{Gravitational radiation reaction and dynamical friction} \label{subsec:staticGWplusDF}

Suppose now that both $\dot r^\mathrm{RR}_2$ and $\dot r^\mathrm{DF}_2$ contribute to the evolution of the binary's separation in Eq.~\eqref{eq:dot_r}.
The amount of time elapsed as the binary inspirals between two orbital radii and the mass captured can be written as integrals over $r_2$ between these radii, and these integrals can be expressed in terms of hypergeometric functions.
For the elapsed time, the integral is
\begin{equation}
    \Delta t = \frac{5 c^5}{64(G m_1)^3} \int^{r_{2,\I}}_{r_{2,\F}} \frac{(r_2)^3 dr_2}{1 + r_2^{11/2-\gamma_\SP}/c_r} \, ,
\end{equation}
where the coefficient $c_r$ was given in~\cite{Kavanagh:2020cfn} as 
\begin{equation}
    c_r = \frac{8 m_1 (Gm_1)^{5/2}}{5\pi c^5 \rho_\SP r_\SP^{\gamma_\SP} \xi \log\Lambda} \, ,
\end{equation}
up to corrections of order $q$.
The elapsed time then can be written as 
\begin{align} \label{eq:Delta_t_DF}
    \Delta t = {} & \Bigg[ \frac{5 c^5}{256(G m_1)^3} r_2^4  \nonumber \\
    & \times {}_2F_1\left(1, b_t, 1+b_t; -\frac{r_2^{11/2-\gamma_\SP}}{c_r}\right) \Bigg ]^{r_{2,\I}}_{r_{2,\F}} \, ,
\end{align}
where $b_t = 8/(11-2\gamma_\SP)$.
The hypergeometric function is bounded between zero and one for positive $r_2$ and $c_r$ (and the values of $\gamma_\SP$ that we consider), so the expression~\eqref{eq:Delta_t_DF} has the form of the fraction of the time to inspiral to zero separation at the initial radius minus that of the final.
Because the hypergeometric function is a decreasing function of radius, this is consistent with the fact that the system will always inspiral more quickly between two given radii when dynamical friction is present than absent.

A similar calculation to solve for the accreted mass shows that 
\begin{align}
    \log \left(\frac{m_{2,\F}}{m_{2,\I}} \right) = {} & \frac{5\pi}{4} \left(\frac{c^2}{Gm_1}\right)^{3/2} \frac{\zeta \rho_\SP r_\SP^{\gamma_\SP}}{m_1} \nonumber \\
    & \times \int^{r_{2,\I}}_{r_{2,\F}} \frac{r_2^{7/2-\gamma_\SP} dr_2}{1 + r_2^{11/2-\gamma_\SP}/c_r} \, . 
\end{align}
The integral can again be evaluated in terms of a hypergeometric function
\begin{align} \label{eq:m_acc_DF_GW}
    \log \left(\frac{m_{2,\F}}{m_{2,\I}} \right) = {} & \Bigg[\frac{5\pi}{4 m_1} \left(\frac{c^2}{Gm_1}\right)^{3/2} \! \! \! \frac{\zeta \rho_\SP r_\SP^{\gamma_\SP} }{9/2-\gamma_\SP} r_2^{9/2-\gamma_\SP} \nonumber \\
    & \times {}_2F_1\left(1, b_m, b_m+1;  -\frac{r_2^{11/2-\gamma_\SP}}{c_r}\right) \Bigg]^{r_{2,\I}}_{r_{2,\F}} \, ,
\end{align}
where $b_m = (9-2\gamma_\SP)/(11-2\gamma_\SP)$.
The function in the exponent for the accreted mass has a similar form to that without dynamical friction, but it is rescaled by a hypergeometric function.
For similar reasons to those described above for the time to inspiral, the amount of accreted mass will be decreased.

The accreted mass in this case is depicted by the gray dotted-dashed ($\zeta=1$) and orange solid ($\zeta=1-\xi$) curves in Fig.~\ref{fig:m_cap_by_m_enc}. 
They show that there is a region of binary primary masses, similar to that of the radiation-reaction only, in which the accreted mass can exceed the enclosed mass.
The deviation of these curves from a power law in $m_1$ with slope $-7/4$ arises from two somewhat competing effects that take place when dynamical friction is included with gravitational radiation reaction.
First, from Eq.~\eqref{eq:Delta_t_DF}, it follows that the binary inspirals from a larger radius in a fixed time interval when dynamical friction is included; this gives the the opportunity to accrete more dark matter over a larger range of radii.
However, dynamical friction speeds up the time to inspiral inward from a given radius, thereby decreasing the amount of time spent at a given radius (and hence the amount of mass accreted at a given radius).
From Eq.~\eqref{eq:m_acc_DF_GW} and Fig.~\ref{fig:m_cap_by_m_enc}, it is possible to deduce that the latter effect is more significant than the former.

Since it was previously demonstrated in~\cite{Kavanagh:2020cfn} that neglecting feedback from dynamical friction on the dark-matter distribution could lead to energy balance being violated significantly during the inspiral, it is perhaps not too surprising that there could be an excess in accreted mass when the halo is assumed to be static.
However, since the dynamical-friction feedback was shown in~\cite{Kavanagh:2020cfn} to produce a large transient depletion of the dark-matter density in the vicinity of the secondary (see also~\cite{Coogan:2021uqv}), it would not be surprising if feedback also had an important effect on the mass captured. 
We describe the effects of including feedback from dynamical friction in the next section.

\section{Evolving dark matter with dynamical-friction feedback} \label{sec:DFfeedback}

We first review the formalism of~\cite{Kavanagh:2020cfn} and the assumptions that enter into this formalism.
We then present results in which we ignore the effects of dark-matter accretion as in~\cite{Kavanagh:2020cfn}, but we consider more massive secondary companions (namely $m_2 = 10 M_\odot)$ than had been studied previously in~\cite{Kavanagh:2020cfn}.
The final part of this section is our treatment of secondary accretion with dynamical-friction feedback, but without attempting to include feedback on the dark-matter distribution from the accretion process.
Dynamical-friction feedback prevents the captured mass from exceeding the initial enclosed mass (for the binaries that we consider), but the ratio of the two masses approaches unity for mass ratios near $q=10^{-2}$ (or $m_1 = 10^3 M_\odot$). 
This suggests that feedback from secondary accretion will be important in this region of parameter space.

\subsection{Review of dynamical-friction feedback} \label{subsec:DFfeedbackReview}

The dynamical-friction feedback introduced in~\cite{Kavanagh:2020cfn} made use of the specific relative energy  
\begin{equation} \label{eq:calE}
    \mathcal E = \frac{Gm_1}r - \frac 12 v^2 \, , 
\end{equation}
where bound orbits are $\mathcal E > 0$, and where we have neglected the potential of the dark-matter spike, which is consistent with our approximation of the Keplerian orbital frequency in Eq.~\eqref{eq:Omega}.\footnote{As a result, we do not introduce the notation $\Psi(r) = \Phi(r) - \Phi_0$, which is used in~\cite{Kavanagh:2020cfn}.}
The distribution function (mass density on phase space) will be assumed to be isotropic in momentum space and spherically symmetric in position space, so that it can be written as just $f(\mathcal E)$ when static, and $f(\mathcal E,t)$ when dynamic.
In the absence of the secondary, a distribution function related through Eddington inversion to the density $\rho_\DM(r)$ in Eq.~\eqref{eq:static_DM} was shown in~\cite{Edwards:2019tzf} to be given by
\begin{equation} \label{eq:static_f}
    f(\mathcal E) = \frac{\gamma_\SP(\gamma_\SP-1)}{(2\pi )^{3/2}} \frac{\Gamma(\gamma_\SP-1)}{\Gamma(\gamma_\SP-1/2)} \left(\frac{r_\SP\mathcal E}{Gm_1}\right)^{\gamma_\SP} \rho_\SP \mathcal E^{-3/2} \, .
\end{equation}
When the secondary is present, it will scatter with dark-matter particles and introduce a time dependence into the dark-matter distribution.

The formalism in~\cite{Kavanagh:2020cfn} relied upon a few key assumptions.
There the dark matter was not modeled on the orbital timescale of the secondary, but only on timescales longer than the orbital period.
During the orbital time, the dark-matter distribution was assumed to equilibrate quickly after scattering, and that on timescales longer than the orbital time the distribution function remains spherically symmetric and isotropic, so that it can be written as a function of just $\mathcal E$ and not also angular momentum.\footnote{It is noted in Ref.~\cite{Kavanagh:2020cfn} that dynamical friction also provides a torque that causes the angular momentum of the binary to decrease.
Thus, through angular momentum conservation, the dark-matter distribution should have some dependence on angular momentum.
However, it was argued in~\cite{Kavanagh:2020cfn} that for the distribution function to develop a strong dependence on angular momentum, the dark-matter particles would need to undergo a large number of scatterings that would tend to unbind the particles from dark-matter spike.
Thus, the distribution of bound dark-matter particles could be approximated reasonably by a distribution function that depends only on the specific energy $\mathcal E$.}
On the longer dissipative timescale, the dark-matter dynamics were modeled by considering the average interactions between the secondary and the dark-matter distribution over an orbital period.
We will also follow similar assumptions when modeling the accretion of dark matter by the secondary.\footnote{For accretion onto the secondary, it was noted in~\cite{Hughes:2018qxz} that the change in the orbital radius $\dot r_2^\mathrm{A}$ could be understood as the change in the orbit that occurs when the angular momentum remains an adiabatic invariant as the secondary's mass increases.
Thus, considerations of angular-momentum balance do not require that angular-momentum dependence of the distribution function change in response to accretion onto the secondary.}

Because $f(\mathcal E,t)$ is the phase-space mass density of particles with a given $\mathcal E$ per volume in position and velocity space, it is convenient to introduce the density of states at each energy, which is given by
\begin{equation}
    g(\mathcal E) = \int d^3 r \int d^3 v \, \delta\boldsymbol( \mathcal E - \mathcal E(r,v)\boldsymbol) \, .
\end{equation}
Using the definition of $\mathcal E$ in Eq.~\eqref{eq:calE}, this can be written as (see, e.g.,~\cite{Kavanagh:2020cfn})
\begin{equation}
    g(\mathcal E) = \sqrt 2 (\pi G m_1)^3 \mathcal E^{-5/2} \, .
\end{equation}
Note that it has units of phase-space volume per energy, so that $f(\mathcal E) g(\mathcal E)$ has units of mass per energy.

The density of states is used to compute the differential scattering rate, per energy change $\Delta \mathcal E$ and per orbital period, of particles with energy $\mathcal E$ to an energy $\mathcal E- \Delta\mathcal E$.
This was given in~\cite{Kavanagh:2020cfn} by
\begin{align}
    \mathcal R_{\mathcal E}(\Delta \mathcal E) = \frac 1{T_2 g(\mathcal E)} \int d^3 r \int d^3 v & \, \delta\boldsymbol( \mathcal E - \mathcal E(r,v)\boldsymbol) \nonumber \\
    & \times \delta\boldsymbol( \Delta\mathcal E(b) - \Delta \mathcal E\boldsymbol) \, .
\end{align}
In the equation above, $b$ is the impact parameter for a scattering event with energy change $\Delta \mathcal E$, and $T_2$ is the orbital period of the secondary.
In a lengthy calculation outlined in Sec.~IV and Appendix~D of~\cite{Kavanagh:2020cfn}, this integral over phase space can be reduced to an integral over a torus with minor radius $b$ and major radius $r_2$.
In the limit that $b/r_2$ is small, the leading-order expression for the integral in this small parameter can be expressed in terms of incomplete elliptic integrals of the second kind, which speeds up the computation of this differential scattering rate. 
We do not make any changes to the calculation in~\cite{Kavanagh:2020cfn} aside from the fact that the notation $\mathcal R_{\mathcal E}(\Delta \mathcal E)$ for the rate that we use is related to the scattering probability of~\cite{Kavanagh:2020cfn} by $P_{\mathcal E}(\Delta \mathcal E)=T_2 \mathcal R_{\mathcal E}(\Delta \mathcal E)$.
Thus, we do not reproduce all the expressions for the impact parameter and scattering rate here.
Finally, it will also be useful to compute the total scattering rate at a given energy, which we denote by
\begin{equation}
    R_{\mathcal E} = \int d\Delta\mathcal E \, \mathcal R_{\mathcal E}(\Delta \mathcal E) \, .
\end{equation}

With these quantities defined, we can write the prescription of~\cite{Kavanagh:2020cfn} for evolving the distribution function due to dynamical-friction feedback.
The basic principle is similar to that of chemical kinetics, where the distribution function takes the place of the concentrations, and scattering rates replace the rate constants.
Specifically, scattering takes away particles with a given energy $\mathcal E$ at the rate $R_{\mathcal E}$ in a way that is proportional to the phase-space density of particles at that energy $f(\mathcal E,t)$.
This leads to an ``outflux'' term of the form $-R_{\mathcal E} f(\mathcal E,t)$.
However, scattering also adds particles at energy $\mathcal E$ by scattering from other energies $\mathcal E - \Delta \mathcal E$ to the energy $\mathcal E$.
This ``influx'' term involves an integral of the form
\begin{equation}
    \int d\Delta\mathcal E \left( \frac{\mathcal E}{\mathcal E - \Delta \mathcal E}\right)^{5/2} \mathcal R_{\mathcal E-\Delta\mathcal E}(\Delta \mathcal E) f(\mathcal E - \Delta\mathcal E,t) \, .
\end{equation}
The first term in the integrand is the ratio of the densities of states at energies $\mathcal E - \Delta\mathcal E$ and $\mathcal E$.
The net influxes and outfluxes then give the following prescription for the  evolution of the distribution function:
\begin{align} \label{eq:fIPDE}
    & \frac{\partial f(\mathcal E,t)}{\partial t} = -R_{\mathcal E} f(\mathcal E,t)   \nonumber \\
    &+ \int d\Delta\mathcal E \left( \frac{\mathcal E}{\mathcal E - \Delta \mathcal E}\right)^{5/2} \mathcal R_{\mathcal E-\Delta\mathcal E}(\Delta \mathcal E) f(\mathcal E - \Delta\mathcal E,t) \, .
\end{align}

There is an implicit dependence on the position of the secondary $r_2(t)$ in the scattering rate $\mathcal R_{\mathcal E}(\Delta \mathcal E)$ (and thus also $R_{\mathcal E}$), because the impact parameter depends upon the radial position of $r_2$.
This implies that the integro-partial-differential equation~\eqref{eq:fIPDE}---or IPDE, for short---is coupled to the ordinary differential equation (ODE) describing the evolution of $r_2$ in Eq.~\eqref{eq:dot_r}.
In addition, because the total mass density in position space is computed via
\begin{equation} \label{eq:rhoDMt}
    \rho_\DM(r,t) = \int d^3v f(\mathcal E,t) \, ,
\end{equation}
the dynamical-friction and the secondary-accretion terms $\dot r_2^\DF$ and $\dot r_2^\mathrm{A}$ are coupled to the IPDE through $\rho_\DM(r_2,t; v<v_2)$ and $\rho_\DM(r_2,t)$, respectively.
Thus, the evolution of $\dot r_2$ and the IPDE must be solved as a coupled IPDE-ODE system.
The \textsc{HaloFeedback} code~\cite{HaloFeedback} implements this procedure to evolve the distribution function and solve the coupled IPDE-ODE system.
We use this code to produce the results in the next subsection.

\subsection{Results without secondary accretion} \label{subsec:DFnoSA}

As a baseline for our comparisons of the effects of secondary accretion on IMRIs with dynamical dark-matter distributions, we evolve a set of five binaries with different mass ratios for which we do not include the effect of the secondary accretion in the evolution equation for $\dot r_2$ (so that the simulations follow the same method as those in~\cite{Kavanagh:2020cfn}).
We do this for two reasons: First, we would like to consider mass ratios closer to one than were simulated in~\cite{Kavanagh:2020cfn} to better compare with the cases treated in~\cite{Yue:2017iwc} for static halos.
Second, although we will treat some of the same mass ratios as in~\cite{Kavanagh:2020cfn}, we will use a larger secondary mass $m_2$ that is more appropriate for a black hole (whereas~\cite{Kavanagh:2020cfn} used a mass appropriate for a neutron star).
Unlike vacuum black-hole binaries, those with dark matter have an additional mass scale (that of the dark matter), which implies that the total mass of the two black holes does not scale out of the problem.
As a result, it is not clear that we can rescale some of the results in~\cite{Kavanagh:2020cfn} to apply to our case with a larger secondary mass.

Specifically, we consider a secondary with mass $m_2=10M_\odot$ and five different primary masses $m_1=10^3$, $3\times10^3$, $10^4$, $3\times10^4$, and $10^5 M_\odot$ (i.e., initial mass ratios of $q=10^{-2}$, $3\times10^{-3}$, $10^{-3}$, $3\times10^{-4}$, and $10^{-4}$).
We evolve the system for an initial dark-matter spike with a power law of $\gamma_\SP=7/3$ and with $\rho_\SP = 226 M_\odot/\mathrm{pc}^3$ for all mass ratios.
We compute the evolution using an initial separation that is three times the separation at which the binary would merge in vacuum in four years ($3 r_\fy$), which is computed assuming an inspiral driven by the Newtonian-order quadrupole formula.
As discussed in further detail in~\cite{Kavanagh:2020cfn}, when the binary starts its inspiral at this separation, the dark-matter distribution at radii smaller than $r_\fy$ is largely unaffected even as the dark-matter distribution reaches a ``steady-state'' configuration during the slow, quasicircular inspiral from $3r_\fy$.
This then makes the dark-matter distribution when the secondary reaches $r_\fy$ consistent with a formation history involving an adiabatic inspiral from much larger radii (something which is not true of the part of the inspiral much closer to $3r_\fy$).
We compute the number of gravitational-wave cycles from the separation of $r_\fy$.
The \textsc{HaloFeedback} code solves the ODE-IPDE system with a maximum time step that is a multiple of the orbital period at a given radius; for the simulations below, we chose this to be 50 orbital periods.\footnote{We performed numerical convergence tests to verify that the binary separation and phase converged at a rate consistent with the second-order method used to solve the IPDE-ODE system using time steps of 50 orbital periods and larger.}
Since the gravitational waves are quadrupolar in the leading, Newtonian approximation, this corresponds to 100 gravitational-wave cycles.
Thus, we will round our expressions for the number of cycles and the amount of dephasing to the nearest 100 cycles here and below.

The number of gravitational-wave cycles and the amount of dephasing from vacuum signals is presented in Table~\ref{tab:DFonly}.
There are several ways to compute these quantities; in terms of the separation $r_2$, the number of cycles can be written as
\begin{equation} \label{eq:Ncycles}
    N_\mathrm{cycles} = \frac 1\pi \int_{r_{2,\I}}^{r_{\ISCO}} \! \Omega \, \dot r_2^{-1} dr_2 \, ,
\end{equation}
where $\Omega$ is the Keplerian orbital frequency given in Eq.~\eqref{eq:Omega}.
We will also find it useful to consider the number of cycles as a function of the starting frequency by mapping the orbital separation to the gravitational-wave frequency using Kepler's third law.
The difference in the number of cycles (the dephasing) is given by
\begin{equation} \label{eq:DeltaNcycles}
    \Delta N_\mathrm{cycles}^{\mathrm{(0-1)}} = N_\mathrm{cycles}^\mathrm{(0)} - N_\mathrm{cycles}^\mathrm{(1)} \, , 
\end{equation}
where $N_\mathrm{cycles}^\mathrm{(0)}$ is the number of cycles in vacuum and $N_\mathrm{cycles}^\mathrm{(1)}$ is the number when dynamical-friction feedback is included.
Because the binary inspirals more quickly with the additional source of energy loss from the binary via dynamical friction, the dephasing $\Delta N_\mathrm{cycles}^{\mathrm{(0-1)}}$ is a positive quantity.

\begin{table}[t!]
    \centering
    \caption{\label{tab:DFonly} \textbf{Number of gravitational-wave cycles with dynamical-friction feedback and dephasing from vacuum binaries}. We use the notation $N_\mathrm{cycles}^\mathrm{(1)}$ for the number when dynamical-friction feedback is included and $\Delta N_\mathrm{cycles}^\mathrm{(0-1)}$ for the dephasing from vacuum systems (difference in number of cycles, from the same starting frequency, between vacuum binaries and those with dark matter when DF feedback is included). The secondary mass is $m_2 = 10 M_\odot$, and the initial dark-matter distribution has $\gamma_\SP=7/3$ and $\rho_\SP = 226 M_\odot/\mathrm{pc}^3$.
    The number of cycles is computed four years from merger, where the merger is defined as when the secondary reaches the ISCO.}
    \begin{tabular}{ccc}
    \hline
    \hline
    $m_1 \, [M_\odot]$ & $N_\mathrm{cycles}^\mathrm{(1)}$ & $\Delta N_\mathrm{cycles}^\mathrm{(0-1)}$ \\
    \hline
    $10^{3}$ & 2,098,000 & 700 \\
    $3\times10^{3}$ & 1,591,500 & 1,700 \\
    $10^{4}$ & 1,174,000 & 4,000 \\
    $3\times 10^{4}$ & 886,900 & 6,200 \\
    $10^{5}$ & 650,300 & 4,300 \\
    \hline
    \hline
    \end{tabular}
\end{table}

First, it is useful to compare the results here for mass ratios of $10^{-4}$ and $10^{-3}$ with those in~\cite{Kavanagh:2020cfn}, which have a lower total mass (and chirp mass).
At a fixed mass ratio and for a fixed time to reach the ISCO, the total number of cycles scales like the chirp mass to the $-5/8$ power (in the Newtonian approximation) which also just goes like the primary mass to the same power.
Thus, it is not too surprising that the total number of cycles at these mass ratios is roughly a factor of $10^{-5/8}\approx 0.24$ times smaller than the corresponding results at the same mass ratio in~\cite{Kavanagh:2020cfn}: With dynamical-friction feedback, the inspiral is driven primarily by gravitational radiation reaction. 

The magnitude of the dephasing $\Delta N_\mathrm{cycles}^{\mathrm{(0-1)}}$ (which is computed from a fixed initial frequency between binaries in vacuum and those with dynamical-friction feedback) is somewhat more subtle to compare, at fixed mass ratio, between the results here and in~\cite{Kavanagh:2020cfn}.
Dynamical friction with feedback is determined by an effective density (discussed in~\cite{Coogan:2021uqv}) evaluated at the location of the secondary and the secondary's mass; this effective density is also a function of the binary separation, binary masses, and the initial dark-matter distribution.
Thus, at fixed secondary mass and for similarly parametrized dark-matter densities, we would need to study the effective density and its scaling with the primary mass; however, we will leave studies of this effective density to future work.
Instead, we will focus primarily on the qualitative similarities in the dephasing.

In particular, we note that the dephasing is not a monotonic function of the mass ratio, but it peaks at a mass ratio between $10^{-4}$ and $10^{-3}$ before decreasing.
This was also observed in~\cite{Kavanagh:2020cfn}, where the explanation for this phenomenon was associated with the increased local depletion of the dark-matter density near the secondary for less-extreme mass ratios (i.e., a lower effective density).
We observe here that this effect becomes even more pronounced at less extreme mass ratios;
thus, although the total number of cycles increases, the amount of dephasing actually decreases, thereby leading to both a smaller absolute and relative effect.
We will next turn to some of the implications of this when we introduce accretion onto the secondary in this evolving dark-matter distribution without incorporating feedback on the distribution from accretion.

\subsection{Secondary accretion with dynamical-friction feedback} \label{subsec:DFplusSAonly}

\begin{figure}[t!]
    \centering
    \includegraphics[width=\columnwidth]{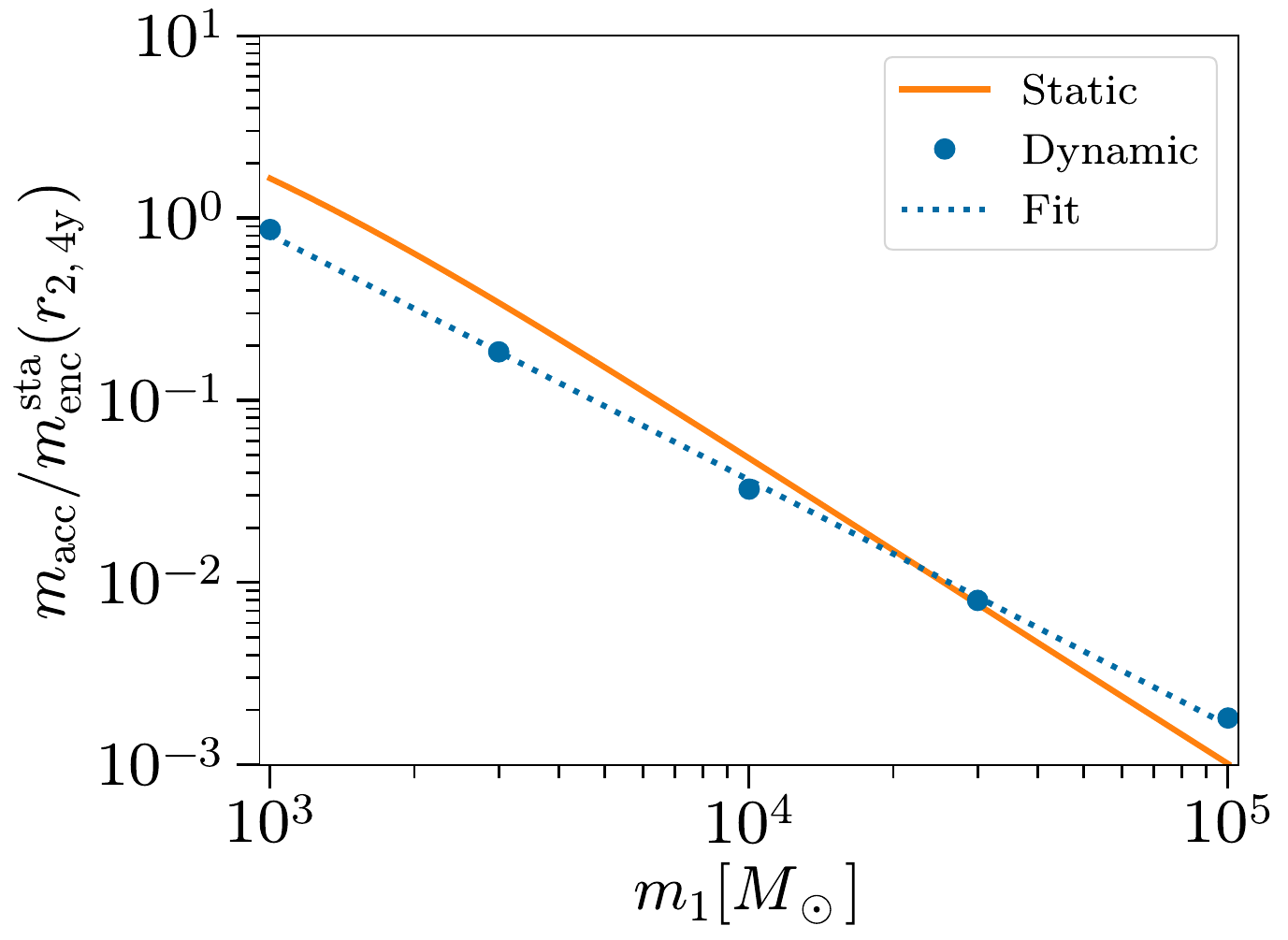}
    \caption{\textbf{The accreted mass $\mathbf{m_{acc}}$ normalized by the enclosed mass $\mathbf{m_{enc}^{sta}(r_{2,4y})}$ of a static dark-matter distribution}.
    The dark matter and binary parameters are chosen as in Fig.~\ref{fig:m_cap_by_m_enc}. 
    The solid orange line is the same $\zeta=1-\xi$ curve in Fig.~\ref{fig:m_cap_by_m_enc} with the secondary's secular evolution driven by radiation reaction and dynamical friction.
    The blue circles are the results of five numerical simulations performed with the \textsc{HaloFeedback} code, and the dotted blue line is a fit to these five points using a single power law.
    Further discussion of the implications of this figure are given in the text of Sec.~\ref{subsec:DFplusSAonly}.}
    \label{fig:m_cap_by_m_enc_DF}
\end{figure}

To understand the effect of dynamical-friction feedback on the accreted mass, we evolve the IMRI with the accretion term in the IMRI equations of motion, so that Eq.~\eqref{eq:dot_r} includes all three terms.
We also compute $m_2$ as a function of time using its evolution equation~\eqref{eq:m2_dot}.
We evolve the five cases described in Sec.~\ref{subsec:DFnoSA} similarly, with the only difference being the additional term in the evolution equation for $\dot r_2$.
The results are summarized in Fig.~\ref{fig:m_cap_by_m_enc_DF} and Table~\ref{tab:DFplusSAonly}.

In Fig.~\ref{fig:m_cap_by_m_enc_DF}, the solid orange curve is the same as the solid orange curve in Fig.~\ref{fig:m_cap_by_m_enc}: namely, the $\zeta=1-\xi$ case with both gravitational waves and dynamical friction driving the evolution of the binary in a static dark-matter distribution.
The five blue circles are the results of the numerical simulations described above for the accreted mass normalized by the enclosed mass.
A power-law least-squares fit to the five data points is shown as the blue dotted curve in Fig.~\ref{fig:m_cap_by_m_enc_DF}.
A single power law with slope $\approx-1.3$ (to two significant figures) is able to capture the qualitative trend in the data. 

A brief comment regarding the convention we use for the enclosed mass for the dynamic halo is in order.
In the dynamic case, we normalized by the enclosed mass within $r_\fy$ when using the initial static dark-matter density to compute the enclosed mass.
For the evolution with dynamical-friction feedback, however, we start the evolution at a distance of $r_{2,\I} = 3 r_\fy$ to produce a dark-matter distribution once the binary reaches a radius of $r_2 = r_\fy$ that is consistent with inspiral from a radius much larger than $r_\fy$.
The dark-matter density when the binary is at $r_\fy$ does differ (significantly in some cases) from the initial power-law distribution in Eq.~\eqref{eq:static_DM} that is used when the binary is at $3r_\fy$ in the dynamic case (or at $r_\fy$ in the static case).
Thus, the mass enclosed using the initial static density, $m_\enc^\sta(r_\fy)$ is different from $m_\enc^\dyn(r_\fy)$, the mass enclosed for the dynamical dark-matter distribution when DF feedback is included.
We choose to normalize $m_\acc$ by $m_\enc^\sta(r_\fy)$ in Fig.~\ref{fig:m_cap_by_m_enc_DF} so that the total accreted mass can be compared more easily between the static and dynamic cases depicted there.
We will also give the ratio $m_\enc^\dyn(r_\fy)/m_\enc^\sta(r_\fy)$ in Table~\ref{tab:DFplusSAonly} which indicates the extent to which dark-matter mass is redistributed once the binary reaches the radius $r_\fy$ in the dynamic case.
It can also be used to determine how efficient the accretion process is in terms of the available amount of enclosed mass that could be accreted.

At the largest primary mass $m_1=10^5 M_\odot$ (or $q=10^{-4}$), the accreted mass in the dynamic case is larger than in the static case for $\zeta=1-\xi$.
This suggests that dynamical friction feedback is not significantly influencing the dark-matter halo and that both dark-matter particles moving more quickly and more slowly than the orbital speed can be accreted by the secondary.
For masses $m_1 \lesssim 3\times 10^4 M_\odot$ (i.e., $q \gtrsim 3\times10^{-3}$), the mass captured in the dynamic case is less than that in the static case, even though in the static case the assumption that $\zeta=1-\xi$ could be understood as representing that only particles moving faster than the secondary's speed are accreted, as compared with dynamical-friction feedback that acts on the more slowly moving particles.
While this may seem surprising, the fact that the number of more slowly moving particles is simply proportional to the total density times $\zeta$ holds for a single-power-law distribution, as in Eq.~\eqref{eq:static_DM}, and not more generally.
Thus, $\zeta \rho_\DM(r_2)$ can overestimate the amount of dark matter at the location of the secondary, particularly when dynamical-friction feedback has a large effect on the dark-matter distribution.

The approximate power-law slope of $\sim -1.3$ in the fit rather than $-7/4$ when radiation-reaction is driving the inspiral of the binary could be understood if the density during the inspiral had a flatter power law than $\gamma_\SP$ in Eq.~\eqref{eq:static_DM} 
(the same calculations in Sec.~\ref{sec:staticAccretion} apply to any power law).
The density would need to follow a power law of roughly $-1.6$ instead of $\gamma_\SP=-7/3$ for this to be the case.
The results for the density in Sec.~\ref{subsec:AllDMdensity} show that the density has a slope that is less steep than the initial dark-matter-spike power law at radii larger than the binary separation, but steeper at smaller radii.
Thus, we did not arrive at a similar, simple explanation for the scaling of the accreted over the enclosed mass with the primary mass when dynamical-friction feedback was included. 

In Table~\ref{tab:DFplusSAonly}, the second column reproduces the gravitational-wave dephasing between vacuum and with dynamical-friction feedback only.
The third column compares the dephasing between dynamical-friction feedback only and including the accretion term in the evolution of $r_2$ without feedback.
(We do not show the total number of cycles in the case, which would be denoted by $N_\mathrm{cycles}^\mathrm{(1A)}$, because the dephasing is typically less than one percent of the total number of cycles.)
We then show the difference $\Delta N_\mathrm{cycles}^\mathrm{(1-1A)}$ between $N_\mathrm{cycles}^\mathrm{(1A)}$ and the number of cycles $N_\mathrm{cycles}^\mathrm{(1)}$ when only dynamical-friction feedback is included (as in Sec.~\ref{subsec:DFnoSA}).
The fourth column $m_\acc/m_\enc^\sta$ contains the same data that appears in the five points in Fig.~\ref{fig:m_cap_by_m_enc_DF}.
The fifth column shows the ratio of the mass enclosed within $r_\fy$ in the dynamic versus the static dark-matter distributions [the latter being known analytically and computed from Eq.~\eqref{eq:mDM_enc}].

\begin{table}[t!]
    \centering
    \caption{\label{tab:DFplusSAonly} \textbf{Number of cycles of dephasing, as well as the normalized accreted and enclosed masses for binaries with dynamical friction and accretion}. The configuration of the dark matter and the binary are the same as in Table~\ref{tab:DFonly}. 
    The second column of dephasing numbers is the same as in Table~\ref{tab:DFonly}, and is reproduced here for ease of comparison.
    The third column of numbers is the dephasing when dynamical-friction feedback is included in both cases, but the effect of accretion on $\dot r_2$ is included in only one case.
    The fourth column contains the same data as the blue points in Fig.~\ref{fig:m_cap_by_m_enc_DF}.
    The final column is the ratio of the dynamical and static enclosed masses, as described further in the text of Sec.~\ref{subsec:DFplusSAonly}.}
    \begin{tabular}{ccccc}
    \hline
    \hline
    $m_1 \, [M_\odot]$ & $\Delta N_\mathrm{cycles}^\mathrm{(0-1)}$ & $\Delta N_\mathrm{cycles}^\mathrm{(1-1A)}$ & $m_\acc/m_\enc^\sta$ & $m_\enc^\dyn/m_\enc^\sta$\\ 
    \hline
    $10^{3}$ & 700 & 3,400 & 0.8640 & 0.645 \\ 
    $3\times10^{3}$ & 1,700 & 1,400 & 0.1843 & 0.674 \\ 
    $10^{4}$ & 4,000 & 600 & 0.0326 & 0.693 \\ 
    $3\times 10^{4}$ & 6,200 & 300 & 0.0080 & 0.729 \\ 
    $10^{5}$ & 4,300 & 200 & 0.0018 & 0.852 \\ 
    \hline
    \hline
    \end{tabular}
\end{table}

Table~\ref{tab:DFplusSAonly} illustrates some clear trends in both the dephasing and mass accreted.
For larger primary masses (above $\sim 10^{4} M_\odot$), the mass accreted is a small fraction of the total mass enclosed within the orbit, and the dephasing induced by accretion is significantly smaller than that due to dynamical friction with feedback.
Nevertheless, the final column (dynamical over static mass enclosed) shows that dynamical-friction feedback does have a nontrivial effect on the distribution of dark matter.
For masses $m_1$ below $\sim 3 \times 10^{3} M_\odot$, the dephasing induced by dynamical friction with feedback and accretion without feedback become comparable, or even several times larger for accretion.
In addition, the amount of accreted mass approaches the enclosed mass as the primary mass approaches $10^{3} M_\odot$.

The results of Fig.~\ref{fig:m_cap_by_m_enc_DF} and Table~\ref{tab:DFplusSAonly} suggest that for smaller $m_1$ (or $q \gtrsim 3 \times 10^{-2}$), the lack of feedback on the distribution function from dark-matter particles being removed leads to a larger amount of dephasing and mass captured than would occur if feedback were included.
We thus turn to introducing a procedure to implement feedback from secondary accretion on the dark-matter distribution in the next part, Sec.~\ref{sec:accretionFeedback}.

\section{Secondary-accretion feedback} \label{sec:accretionFeedback}

In the previous section, we found that with dynamical-friction feedback the amount of mass accreted remains less than the enclosed mass in the cases we have simulated (but the two masses could be nearly equal).
In addition, for the systems in which the largest fraction of the enclosed mass was accreted, the effect of secondary accretion on the evolution of the orbital phase exceeded that of dynamical friction with feedback on the dark matter. 
This result seemed surprising given the higher post-Newtonian nature of the secondary-accretion process, and it suggested that we need a procedure to remove the accreted dark-matter mass from the distribution function so as avoid these scenarios that lead to unreasonably large dark-matter secondary accretion.
To do so, it will then be necessary to evolve the dark-matter distribution in response to the removal of dark-matter particles from the distribution function for particles with orbits that fall within the accretion cross section of the secondary.
We discuss a procedure to implement this removal process in this section.

\subsection{Formalism for secondary-accretion feedback} \label{subsec:SAformalism}

In this part, we derive how the accretion of dark matter modifies the evolution of the distribution function.
The final result is relatively simple: only an additional term of the form $-R_{\mathcal E}^\acc f(\mathcal E,t)$ must be added to Eq.~\eqref{eq:fIPDE} for the distribution function.
Namely, secondary accretion simply removes particles from $f(\mathcal E, t)$ at each energy by an energy-dependent rate $R_{\mathcal E}^\acc$; this causes the magnitude of the distribution function to decrease at all the energies for which $R_{\mathcal E}^\acc$ is nonzero.
This mass loss must be balanced by an increase in mass of the secondary; our prescription for accretion feedback is consistent with the mass accretion rate in Eq.~\eqref{eq:m2_dot}.

\subsubsection{Derivation of secondary-accretion feedback rate}

The procedure that we use is qualitatively similar to that of the dynamical-friction feedback on the dark-matter spike.
We make similar assumptions to those used in dynamical-friction feedback, in particular with regard to the quick equilibration on the orbital timescale (which is used to justify maintaining spherical symmetry on the longer dissipative timescale).
We then will compute the total rate of dark-matter accreted per orbital period at each specific relative energy $\mathcal E$. 

We compute this per-orbit rate of accretion to be 
\begin{equation}
    R^\acc_{\mathcal E} = \frac 1{T_2 g(\mathcal E)} \int_{\mathbf{r} \in T^2} d^3 r \int d^3 v \, \delta\boldsymbol( \mathcal E - \mathcal E(r,v)\boldsymbol) \, .
\end{equation}
The domain of the integral over position, denoted by $\mathbf{r} \in T^2$, indicates that it should take place over the interior of a torus of major radius $r_2$ and of minor radius $b_\acc = \sqrt{\sigma(v_2)/\pi}$.
The integral over $v$ can be evaluated using the properties of the delta function, but some care must be taken when doing this.
The result of this integration is a square root, which must be positive for the rate to be real.
Because we consider only bound orbits, then the rate is restricted to values of $\mathcal E$ that satisfy $\mathcal E \in [0, Gm_1/r]$, for values of $r$ where there are at least some values that lie in the torus.
However, for simplicity we will make the further assumption that if $\mathcal E$ satisfies $\mathcal E > Gm_1/r$ for any value of $r \in [r_2-b_\acc,r_2+b_\acc]$, then the rate for this energy is zero.
In this approximation, the result of integrating over velocities is
\begin{align} \label{eq:RaccE1}
    & R^\acc_{\mathcal E} = \nonumber \\
    & \begin{cases}
    \dfrac{4\pi\sqrt 2}{T_2 g(\mathcal E)} \displaystyle \int_{\mathbf{r} \in T^2} \! \! d^3 r \, \sqrt{\dfrac{Gm_1}{r} - \mathcal E} & \mbox{for} \ \ \mathcal E < \dfrac{Gm_1}{r_2+b_\acc} \\
    0 & \mbox{for} \ \ \mathcal E \geq \dfrac{Gm_1}{r_2+b_\acc}
    \, .
    \end{cases}
\end{align}

We now argue that this approximation will have a small effect on the final expression for the rate.
To do so, it is useful to first note that the ratio of the inner radius to the outer radius of the torus can be written as 
\begin{equation}
    \frac{b_\acc}{r_2} = \frac{4Gm_2}{c r_2 v_2} = 4 q \frac{v_2}{c} \, ,
\end{equation}
where $Gm_1/r_2 = (v_2)^2$ was used in the second equality.
Thus, even as $v_2$ becomes relativistic, the ratio of the radii is always a small quantity of order $q$. 
This implies that we can write $G m_1/r$ on the domain of integration of the torus as 
\begin{equation}
    \frac{G m_1}{r} \approx \frac{G m_1}{r_2}[1+O(q)] \, 
\end{equation}
and we can neglect the $O(q)$ terms. 
Similarly, this means that when we consider the integral in Eq.~\eqref{eq:RaccE1} for energies $\mathcal E < Gm_1/(r_2 + b_\acc)$, then to good approximation, we can replace $\mathcal E < Gm_1/(r_2 + b_\acc)$ with $\mathcal E < Gm_1/r_2$.
This also shows that our approximation for the value of the energy at which the rate vanishes had errors of order $q$; however, we have frequently worked to leading order in $q$ throughout this paper.

With $G m_1/r \approx G m_1/r_2$, the integrand can be treated as constant on the torus, so the integral in Eq.~\eqref{eq:RaccE1} reduces to the integrand evaluated at $r_2$ times the volume of the torus.
As a result, the secondary-accretion rate for particles of energy $\mathcal E$ will be given by
\begin{equation} \label{eq:RcaptEcases}
    R^\acc_{\mathcal E} = 
    \begin{cases} 
    8\pi^2\sqrt 2\dfrac{r_2 \sigma(v_2)}{T_2 g(\mathcal E)}  \sqrt{\dfrac{Gm_1}{r_2} - \mathcal E} & \mbox{for} \ \ \mathcal E < \dfrac{Gm_1}{r_2} \\
    0 & \mbox{for} \ \ \mathcal E \geq \dfrac{Gm_1}{r_2} \, ,
    \end{cases}
\end{equation}
where we still consider only bound orbits with $\mathcal E > 0$.

Using the results in~\cite{Kavanagh:2020cfn}, one can show that $R^\acc_{\mathcal E}/R_{\mathcal E}$ scales as $(v_2/c)^2$, so the secondary-accretion rate is one PN order higher than the the rate of feedback from dynamical friction.
This is similar to the fact that the dissipative effects in the equations of motion for the IMRI are one PN order higher for dark-matter accretion than they are for dynamical friction.
However, unlike dynamical friction, which preferentially transfers energy to dark matter particles that are moving more slowly than the orbital speed $v_2$, dark-matter accretion affects both the more slowly moving and the more rapidly moving particles.
Consequently, while secondary accretion will have a weaker effect on the more slowly moving dark matter particles at a given $r$, it will have a leading-order effect on the distribution of dark matter for the more rapidly moving particles at a given $r$.

\subsubsection{Evolution equations with secondary feedback and mass conservation}

Next, we discuss how secondary feedback affects the coupled IPDE-ODE system that describes the evolution of the IMRI and the surrounding dark matter.
Secondary accretion adds one new term to the IPDE in Eq.~\eqref{eq:fIPDE} of the form $ R_{\mathcal E}^\acc f(\mathcal E,t)$, so that the IPDE can be written as
\begin{align} \label{eq:fIPDEcap}
    & \frac{\partial f(\mathcal E,t)}{\partial t} = -(R_{\mathcal E} + R_{\mathcal E}^\acc ) f(\mathcal E,t) \nonumber \\
    &+ \int d\Delta\mathcal E \left( \frac{\mathcal E}{\mathcal E - \Delta \mathcal E}\right)^{5/2} \mathcal R_{\mathcal E-\Delta\mathcal E}(\Delta \mathcal E) f(\mathcal E - \Delta\mathcal E,t) \, .
\end{align}
A key difference between the dynamical-friction and secondary-accretion feedback is that secondary-accretion feedback removes particles from the distribution function without replacing them (they fall into the secondary black hole), whereas dynamical-friction feedback redistributes particles with slower speeds to those with greater speeds.
Thus, dynamical-friction feedback largely conserves mass in the distribution function (aside from some particles scattering onto unbound orbits), whereas secondary-accretion feedback causes the total mass of the dark-matter distribution to decrease; however, the loss of mass from the dark-matter distribution should be balanced precisely by an increase in mass of the secondary.

It is not obvious, \textit{a priori}, that the formalism for accretion feedback introduced above will lead to a loss of mass from the dark-matter distribution that is consistent with the increase in mass of the secondary given in Eq.~\eqref{eq:m2_dot}.
We can prove that the two are consistent in a few lines, however.
To do so, we integrate over phase space [using the density of states $g(\mathcal E)$] the term $-R^\acc_\mathcal{E} f(\mathcal E, t)$ that governs the loss of dark-matter mass from accretion feedback:
\begin{equation}
    \frac{d m_\DM}{dt} = - \int R^\acc_\mathcal{E} f(\mathcal E, t) g(\mathcal E) d\mathcal E \, .
\end{equation}
This reduces, for $R^\acc_\mathcal{E}$ in Eq.~\eqref{eq:RcaptEcases}, to
\begin{equation} \label{eq:dmDMdtInt}
    \frac{d m_\DM}{dt} = - 4\pi v_2 \sigma(v_2) \int_0^{\tfrac{Gm_1}{r_2}} \! \! \! d\mathcal E \, f(\mathcal E, t) \sqrt{2\left(\frac{Gm_1}{r_2} - \mathcal E\right)}  \, .
\end{equation}
The expression~\eqref{eq:dmDMdtInt} has the same dependence on $v_2$ and on the cross section $\sigma(v_2)$ as the evolution of $m_2$ in Eq.~\eqref{eq:m2_dot}.
The integral can be shown to be proportional to the density at $r_2$ by writing the energy at $r_2$ as $\mathcal E = \mathcal E(r_2, v)$ and using $d\mathcal E = v dv$ at fixed $r_2$.
The square root in the integrand reduces to the speed $v$.
Writing $f\boldsymbol(\mathcal E(r_2,v), t\boldsymbol) = f(r_2, v, t)$, then the integral reduces to
\begin{equation}
    \frac{d m_\DM}{dt} = - v_2 \sigma(v_2) \int_0^{v_\mathrm{max}} \! \! f(r_2, v, t) 4\pi v^2 dv  \, ,
\end{equation}
where $v_\mathrm{max} = \sqrt{2Gm_1/r_2}$ is the maximum bound velocity at $r_2$.
The integral is written now in a form in which it is more clearly equal to $\rho_\DM(r_2,t)$; this shows that the rate of mass loss from the distribution function is given by 
\begin{equation} \label{eq:dmDMdt}
    \frac{d m_\DM}{dt} = - v_2 \sigma(v_2) \rho_\DM(r_2, t) \, .
\end{equation}
Thus, the accretion feedback rate is consistent with the accretion rate onto the secondary, Eq.~\eqref{eq:m2_dot}.
This implies that secondary-accretion feedback as implemented here conserves the combined mass of dark matter and the secondary:
\begin{equation} \label{eq:dotm2SA}
    \dot m_2 = - \frac{dm_\DM}{dt} \, .
\end{equation}
We can then continue to compute the accretion term $\dot r_2^\mathrm{A}$ in the equation of motion for $\dot r_2$ as before.
As a first study of the secondary-accretion-feedback term, we investigate how it behaves in isolation (without dynamical-friction and its feedback) in the next part.

\subsection{Results with secondary-accretion feedback but without dynamical-friction feedback} \label{subsec:SAonly}

To help understand the properties of secondary-accretion feedback on the dark-matter distribution, we first consider a simpler test case of how accretion influences the distribution without dynamical-friction feedback. 
In this case, the IPDE for the evolution of the distribution function reduces to a standard PDE:
\begin{equation} \label{eq:PDEcaptOnly}
    \frac{\partial f}{\partial t} = -R_\mathcal{E}^\acc f(\mathcal E, t) \, .
\end{equation}
Since $R_\mathcal{E}^\acc$ depends on the position $r_2$, the PDE is coupled to the ODE for $r_2$ (and vice versa).
This coupling makes it more challenging to find general analytical solutions, but there are some solutions that can be found when additional approximations are made.
These simpler scenarios can provide some intuition about the secondary-accretion process.

First, when the secondary is held at a fixed location, the secondary-accretion rate $R_\mathcal{E}^\acc$ is no longer time dependent.
We can then integrate Eq.~\eqref{eq:PDEcaptOnly} directly to write it in the form
\begin{equation}
    f(t,\mathcal E) = f(\mathcal E) \exp\left(-R_\mathcal{E}^\acc t\right) \, ,
\end{equation}
where $f(\mathcal E)$ is the initial value of the distribution function at time $t=0$.
Accretion then produces an exponential decay of the distribution function at each specific energy at the rate $R_{\mathcal E}^\acc$, for energies for which the rate is nonzero. 
The expression for the rate in Eq.~\eqref{eq:RcaptEcases} has the properties that it goes to zero at both $\mathcal E=0$ and $\mathcal E = Gm_1/r_2$, and it peaks at an energy equal to $(5/6)(Gm_1/r_2)$.
The distribution function then gets depleted most strongly around this value of the energy.
The dark-matter density, being an integral over velocity space of the distribution function, has a more nontrivial profile as a function of position from the accretion process (as we will show in more detail below when we do not keep the secondary's location fixed).

Second, in the approximation in which the binary's separation $r_2$ evolves under the effect of gravitational radiation reaction only, the distribution function as a function of $\mathcal E$ and time $t$ can again be determined analytically.
In this case, it is more convenient to use the chain rule to write the differential equation as
\begin{equation}
    \frac{\partial f}{\partial r_2} = - \frac{R_\mathcal{E}^\acc}{\dot r_2} f(\mathcal E, r_2) \, .
\end{equation}
For the energies for which $R_\mathcal{E}^\acc$ is nonzero, the product $R_\mathcal{E}^\acc (\dot r_2)^{-1}$ can be written as 
\begin{equation}
    \frac{R_\mathcal{E}^\acc}{\dot r_2} = 
    \frac{5q c^3 r_2^{7/2} \mathcal E^{5/2}}{\pi (G m_1)^{9/2}} \sqrt{\frac{Gm_1}{r_2} - \mathcal E} \, .
\end{equation}
If we then define the ``energy radius'' by
\begin{equation}
    r_\mathcal{E} = \frac{G m_1}{\mathcal E} 
\end{equation}
and the normalized (dimensionless) radius by
\begin{equation}
    r_{2/\mathcal E} = \frac{r_2}{r_\mathcal{E}} \,
\end{equation}
then the distribution function has the following reasonably simple form in terms of the changes in the initial and final normalized radii $r_{2/\mathcal E}$.
\begin{align} \label{eq:f_of_r2_E}
    \log\frac{f(r_{2,\F},\mathcal E)}{f(r_{2,\I},\mathcal E)} = {} & \Delta \left[\frac{  2 q c^3}{63\pi \mathcal E^{3/2}} (1-r_{2/\mathcal E})^{3/2} \theta\left(\frac{Gm_1}{r_2} - \mathcal E\right) \right. \nonumber \\
    & \times \theta(\mathcal E) (35 r_{2/\mathcal E}^3 + 30 r_{2/\mathcal E}^2 + 24 r_{2/\mathcal E} + 16) \Bigg] \, .
\end{align}
The $\Delta$ means that the difference of the expression at the final and initial radii $r_{2/\mathcal E}$ should be taken.
The two unit step functions $\theta(x)$ are required to set the argument of the exponential to zero when the secondary-accretion rate goes to zero.
The expression~\eqref{eq:f_of_r2_E} is implicitly a function of an interval of time, $\Delta t$, because when $r_2$ evolves because of radiation reaction, the final value of $r_2$ is given by
\begin{equation}
    r_{2,\F} = \left(r_{2,\I}^4 - \frac{256 q(Gm_1)^3}{5c^5} \Delta t \right)^{1/4} \, .
\end{equation}

The analytical expression for $f(r_2,\mathcal E)$ in Eq.~\eqref{eq:f_of_r2_E} allows one to see that as the secondary inspirals between an initial radius $r_{2,\I}$ and the ISCO radius, $r_\mathrm{ISCO}$, dark-matter particles with smaller $\mathcal E$ are much more efficiently accreted than those with larger $\mathcal E$.
This likely occurs because secondary-accretion feedback occurs only up to the energy $Gm_1/r_2$ but down to an energy of $\mathcal E = 0$.
When the secondary evolves as a result of radiation reaction, it spends a larger number of orbital periods at larger separations, which allows it to accrete more dark matter with specific energies closer to zero during the inspiral.
Only late in the evolution do the particles with larger $\mathcal E$ become accessible to accretion feedback.

\begin{figure*}[t!]
    \centering
    \includegraphics[width=0.49\textwidth]{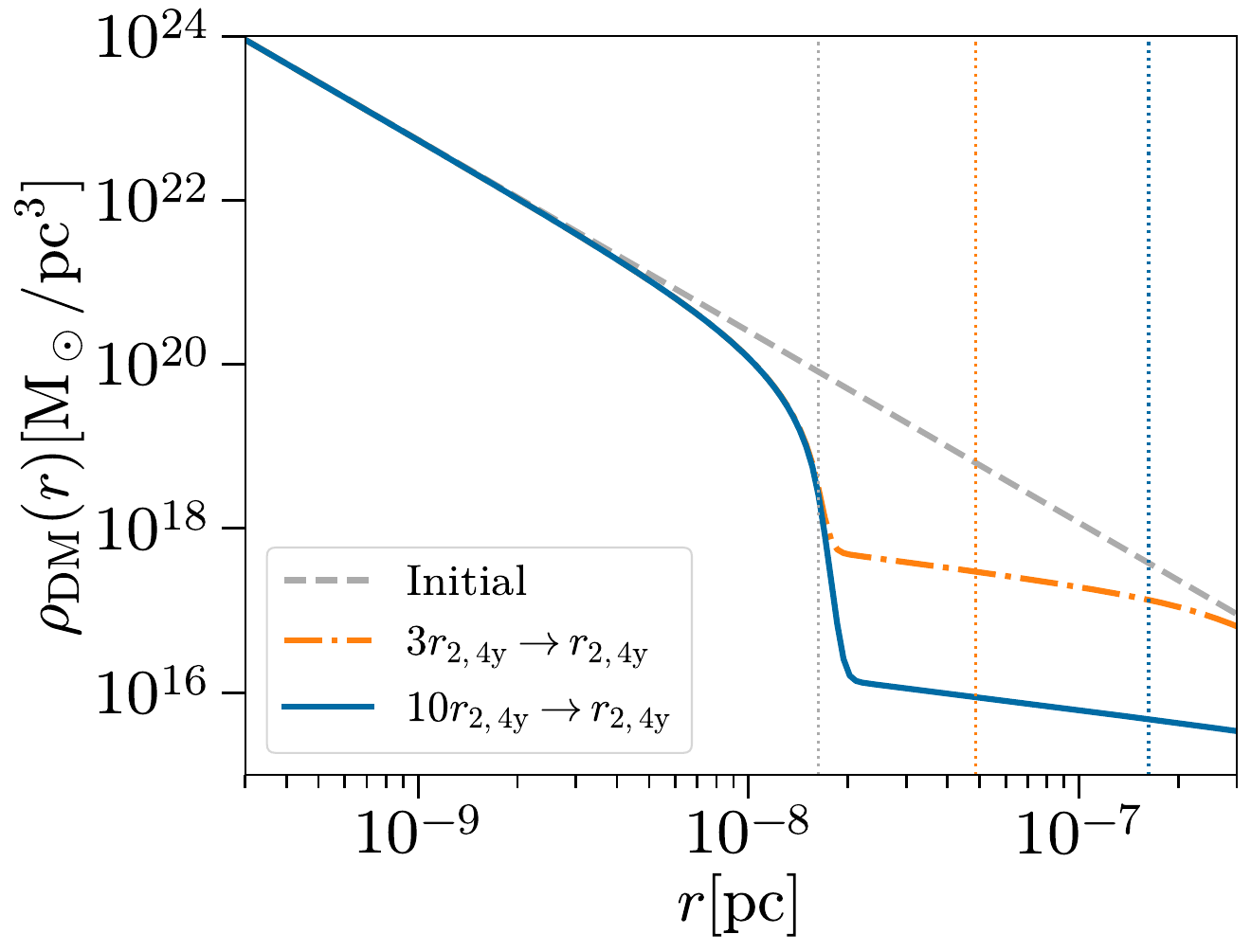} \
    \includegraphics[width=0.49\textwidth]{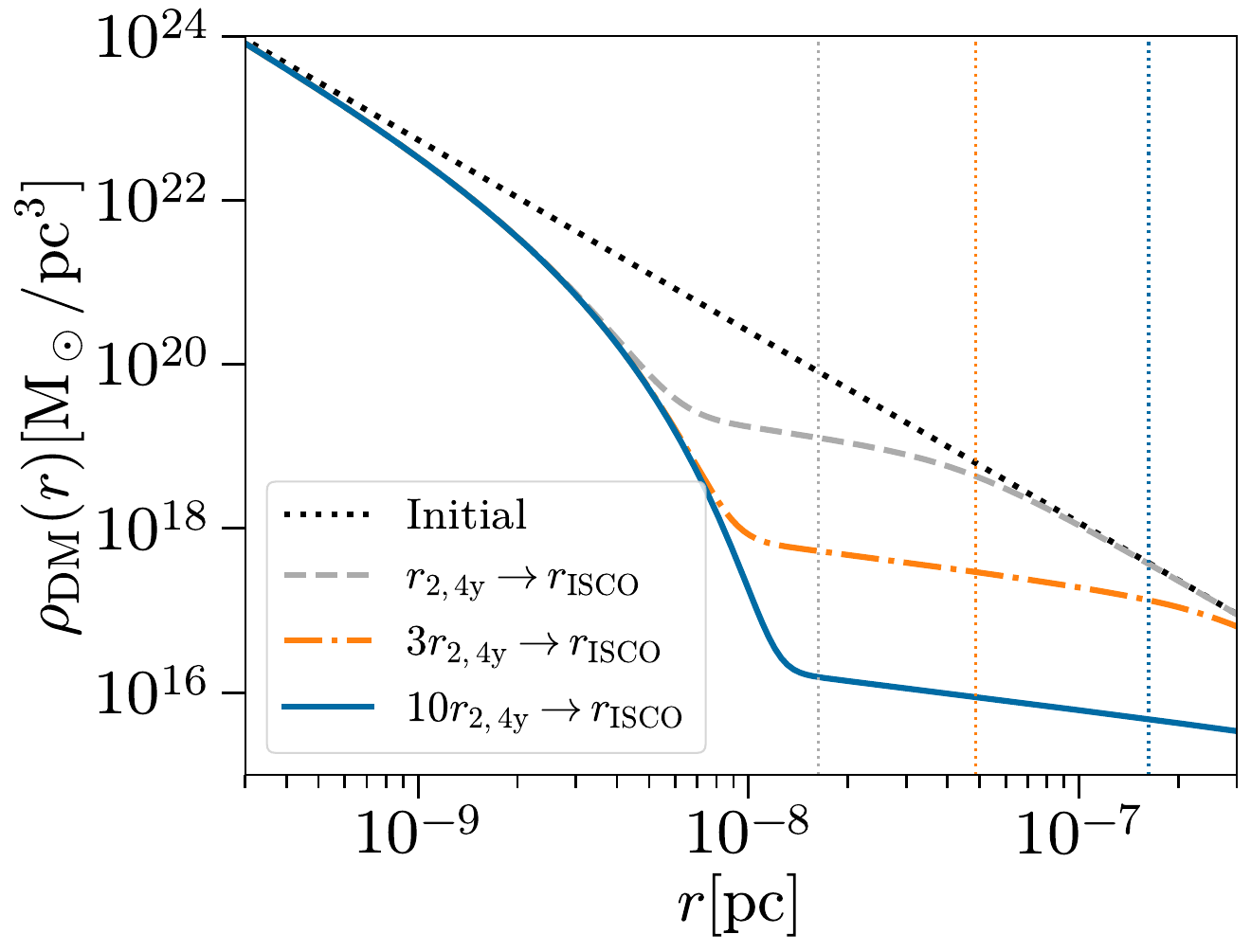}
    \caption{\textbf{Dark-matter density during and following the inspiral for different initial separations assuming no dynamical-friction feedback and orbital evolution governed by radiation reaction only}.
    In both panels, the binary began as two black holes with masses $m_1=10^3M_\odot$ and $m_2=10M_\odot$.
    The curves labeled ``initial'' are the initial dark-matter density in Eq.~\eqref{eq:static_DM} for $\rho_\SP=226M_\odot/\mathrm{pc}^3$ and $\gamma_\SP=7/3$.
    The radius $r_\fy$ is shown with a vertical light-gray dotted line and the radii $3r_\fy$ and $10r_\fy$ are the dotted orange and blue vertical lines, respectively.
    In the left panel, the two other curves of different colors and line styles (dashed-dotted orange and solid blue) show the densities when the binary reaches the radius $r_\fy$, after having inspiraled from different starting radii ($3r_\fy$ and $10r_\fy$ for orange and blue, respectively).
    This illustrates how the density at radii larger than $r_\fy$ depends strongly on the initial conditions, whereas it does not for radii smaller than $r_\fy$, when the initial radius is larger than $3r_\fy$.
    The right panel shows the density after the secondary reaches the ISCO for the three initial separations of $r_\fy$, $3r_\fy$ and $10r_\fy$ (the dashed gray, dotted-dashed orange, and solid blue curves, respectively).
    The figure is discussed further in the text of Sec.~\ref{subsec:SAonly}.}
    \label{fig:DM_capt_only}
\end{figure*}

We also compute the density $\rho_\DM(r)$ by numerically integrating $f(r_2, \mathcal E)$ over all velocities.
The several different dark-matter densities at different initial and final radii are shown in the left and right panels of Fig.~\ref{fig:DM_capt_only} for a binary with masses $m_1=10^3M_\odot$ and $m_2=10 M_\odot$.
The curves labeled ``initial'' correspond to the static, power-law dark-matter distribution in Eq.~\eqref{eq:static_DM} for $\rho_\SP=226M_\odot/\mathrm{pc}^3$ and $\gamma_\SP=7/3$.
Three different initial conditions for the binary's separation $r_\fy$, $3r_\fy$, and $10 r_\fy$ are considered on the right and just the larger two are treated on the left.
As before, $r_\fy$ is the radius for the binary to inspiral to the ISCO under the influence of only gravitational radiation reaction.
The solid blue curves show the dark-matter distribution after the binary evolves from $r_{2,\I} = 10 r_\fy$ with $f(r_{2,\I},\mathcal E)$ given by Eq.~\eqref{eq:static_f} to $r_{2,\F} = r_\fy$ (on the left) and $r_{2,\F} = r_\ISCO$ (on the right).
The dotted-dashed orange curves are the analogous densities for $r_{2,\I} = 3 r_\fy$.
The dashed gray curve in the right panel shows the density when $r_{2,\I} = r_\fy$ and $r_{2,\F} = r_\ISCO$.
One can also interpret the curves in the right panel as being the result of evolving the curves with the corresponding line styles in the left panel from the same initial separation of $r_{2,\I}=r_\fy$ to $r_{2,\F}=r_\ISCO$.

This latter interpretation of the corresponding curves in the two panels is useful for understanding what are suitable initial starting radii that lead to a robust evolution of the binary and to what extent the final dark-matter distribution is influenced by the choice of initial data (similar questions were considered in~\cite{Kavanagh:2020cfn} when treating dynamical-friction feedback only).
Since for initial radii with $r_\I \gtrsim 3r_\fy$ the dark-matter distribution when the binary reaches $r_\fy$ is nearly identical at radii with $r < r_\fy$ (and since the evolution of the mass $\dot m_2^\acc$ in Eq.~\eqref{eq:m2_dot} depends on just the local density of dark matter), then the calculations of the accreted mass $m_\acc(r_\fy)$ are not too strongly dependent on the initial radius for $r_\I \gtrsim 3r_\fy$.
The same is not true of the dark matter distribution after the secondary reaches the ISCO.
Comparing the orange and blue curves, the two agree in a region of $r \lesssim r_\fy/2$.
Thus, there is a smaller range of radii over which the final dark-matter distribution is insensitive to the choice of initial conditions.
This should be kept in mind when interpreting Fig.~\ref{fig:DM_capt_only}.

Nevertheless, in these calculations that neglect dynamical friction, secondary accretion has a very large effect on the final dark-matter distribution, particularly at larger radii, where the secondary undergoes more orbits before inspiraling to smaller radii.
The effect on the dark matter near the ISCO (the left-side of the plots) is much smaller, because radiation reaction causes the binary to inspiral very rapidly there. 
Unlike dynamical friction, which has a larger transient effect on the particles moving more slowly than the local orbital speed, accretion by the secondary can cause a more significant lasting change to the dark-matter distribution.
Moreover, Fig.~\ref{fig:DM_capt_only} suggests that to model accurately the final distribution of dark matter, it is necessary to know the correct initial conditions of the binary.\footnote{The case of $r_{2,\I}=r_\fy$, for example, corresponds to a scenario in which the binary ``materialized'' at the radius $r_\fy$ in the initial density profile in Eq.~\eqref{eq:static_DM} precisely four years from merging without migrating in from some larger separation.
Similar statements hold for the other separations $r_{2,\I}$ that are multiples of $r_\fy$.\label{fn:formation}}

The results shown in Fig.~\ref{fig:DM_capt_only} should be interpreted with some caution, however.
First, as elaborated on in Footnote~\ref{fn:formation}, the different curves assume an improbable formation scenario in which the binary appears at the initial radius $r_{2,\I}$ without having inspiraled or been captured at a larger radius.
Second, the results neglect feedback from dynamical friction, which we expect to be more efficient, because of its lower post-Newtonian order in the evolution equation for $r_2$.
In particular, this suggests that the feedback from dynamical friction could redistribute particles away from the locations in phase space where they are most efficiently accreted by the secondary.
Thus, the results in Fig.~\ref{fig:DM_capt_only} are likely overestimates of the influence of secondary accretion on the distribution of dark matter after the merger.

As a result, we do not present results for $m_\acc/m_\enc^\sta$ in this subsection, because without dynamical-friction feedback, the accreted mass in this case will certainly be greater than when both types of feedback are included. 
However, we will show the results for $m_\acc/m_\enc^\sta$ in Sec.~\ref{subsec:AllGWmacc} to help explain some of the qualitative features of this ratio for different primary masses.

We now turn to the self-consistent modeling of the binary and dark matter including both dynamical-friction and secondary-accretion feedback in the following section.

\section{Results with both types of feedback}
\label{sec:AllFeedback}

We implement the new secondary-accretion feedback term written in Eq.~\eqref{eq:fIPDEcap} by adding it to the \textsc{HaloFeedback} code.
This allows us to solve the full evolution equations in~\eqref{eq:fIPDEcap} with both dynamical-friction and secondary-accretion feedback, when coupled to the ODE with all terms in the evolution equation for $\dot r_2$ and to the evolution for the secondary's mass, $\dot m_2$.
In the subsections below, we focus on the impact of secondary-accretion feedback in this context: in particular, how it changes the number of orbital (or similarly, gravitational-wave) cycles during the merger, influences the accreted mass onto the secondary, and affects the dark-matter distribution during (and after) the inspiral.

\subsection{Gravitational-wave dephasing and accreted mass}
\label{subsec:AllGWmacc}

To help compare with the simulations in Tables~\ref{tab:DFonly} and~\ref{tab:DFplusSAonly}, we again evolve the same five primary and secondary masses as in these tables, with the same initial dark-matter density $\rho_\SP = 226 M_\odot/\mathrm{pc}^3$ and $\gamma_\SP = 7/3$.
We show the results in Table~\ref{tab:DFplusSAfeed}.
The second column is the dephasing between simulations with both DF and SA feedback (where the number of gravitational-wave cycles with both types of feedback is $N_\mathrm{cycles}^\mathrm{(2)}$) and simulations with only dynamical-friction feedback.
We denote this by $\Delta N_\mathrm{cycles}^\mathrm{(1-2)}$.
The third column shows the dephasing between calculations with both types of feedback and DF feedback with accretion onto the secondary, but without feedback from accretion (denoted $\Delta N_\mathrm{cycles}^\mathrm{(2-1A)}$).
The final two columns are the accreted mass normalized by the enclosed mass in the initial (static) halo, as well as the ratio of the mass enclosed in the dynamical halo at a binary separation of $r_\fy$ to the static enclosed mass (for an initial separation of $3r_\fy$, as in the dynamical-friction feedback only simulations).

\begin{table}[t!]
    \centering
    \caption{\label{tab:DFplusSAfeed} \textbf{Two cases of dephasing, as well as normalized accreted and enclosed masses for binaries including dynamical friction and accretion effects}. The configuration of the dark matter and the binary are the same as in Table~\ref{tab:DFonly}. 
    The second and third columns show dephasing in two cases.
    The second is the dephasing between simulations with both types of feedback and with only dynamical friction feedback.
    The third compares both types of feedback to a case with accretion without feedback and dynamical-friction with feedback.
    The fourth and fifth columns are the analogs of the same columns in Table~\ref{tab:DFplusSAonly}, but now the accreted and dynamical enclosed masses are computed in simulations with both types of feedback.
    The interpretation of these numbers is given in the text of Sec.~\ref{subsec:AllGWmacc}.}
    \begin{tabular}{ccccc}
    \hline
    \hline
    $m_1 [M_\odot]$ & $\Delta N_\mathrm{cycles}^\mathrm{(1-2)}$ & $\Delta N_\mathrm{cycles}^\mathrm{(2-1A)}$ & $m_\acc/m_\enc^\sta$ & $m_\enc^\dyn/m_\enc^\sta$\\
    \hline
    $10^{3}$ & 500 & 2,100 & 0.2461 & 0.604 \\
    $3\times10^{3}$ & 800 & 700 & 0.1167 & 0.641 \\
    $10^{4}$ & 500 & 100 & 0.0302 & 0.686 \\
    $3\times 10^{4}$ & 200 & 100 & 0.0079 & 0.727 \\
    $10^{5}$ & 200 & 0 & 0.0018 & 0.851 \\
    \hline
    \hline
    \end{tabular}
\end{table}

A few comments are in order about the results in Table~\ref{tab:DFplusSAfeed}, especially with regards to how the numbers here compare with those in Table~\ref{tab:DFplusSAonly}.
First, comparing the second and third columns, it is clear that feedback has a relatively small effect for primary masses larger than $10^4 M_\odot$ ($q \lesssim 10^{-3})$.
However, for smaller $m_1$, neglecting feedback can lead to an overestimate of the amount of dephasing by a multiplicative factor of a few.
Another noteworthy feature of adding feedback is that the amount of dephasing peaks at a mass ratio of around $3\times10^{-3}$, similarly to how the simulations with DF feedback only have a maximum dephasing at a mass ratio about ten times smaller.
One way to understand this behavior relies on an explanation similar to that given to explain the peak dephasing that occurs with DF feedback.
As $m_1$ decreases, accretion becomes more efficient and feedback becomes stronger.
At some point, it becomes so strong that there is a significant depletion of the dark-matter density, which causes the amount of dephasing to decrease (since the effects of accretion on $\dot r_2$ depend linearly on the density in the differential equation).
Likely due to the higher post-Newtonian order of accretion, the peak mass ratio for the dephasing takes place at a less extreme value than it does with dynamical friction.

Comparing the values of the accreted and enclosed masses in Tables~\ref{tab:DFplusSAonly} and~\ref{tab:DFplusSAfeed}, respectively, one can similarly see that feedback does have a relatively small effect for $m_1$ greater than $\sim 10^{4} M_\odot$.
At larger mass ratios, the total accreted mass is now well below one, but still a significant fraction of the total enclosed mass.
Thus, feedback plays an important role in enforcing the conservation of mass and in producing more reliable estimates of the amount of gravitational-wave dephasing. 

\begin{figure}[t!]
    \centering
    \includegraphics[width=\columnwidth]{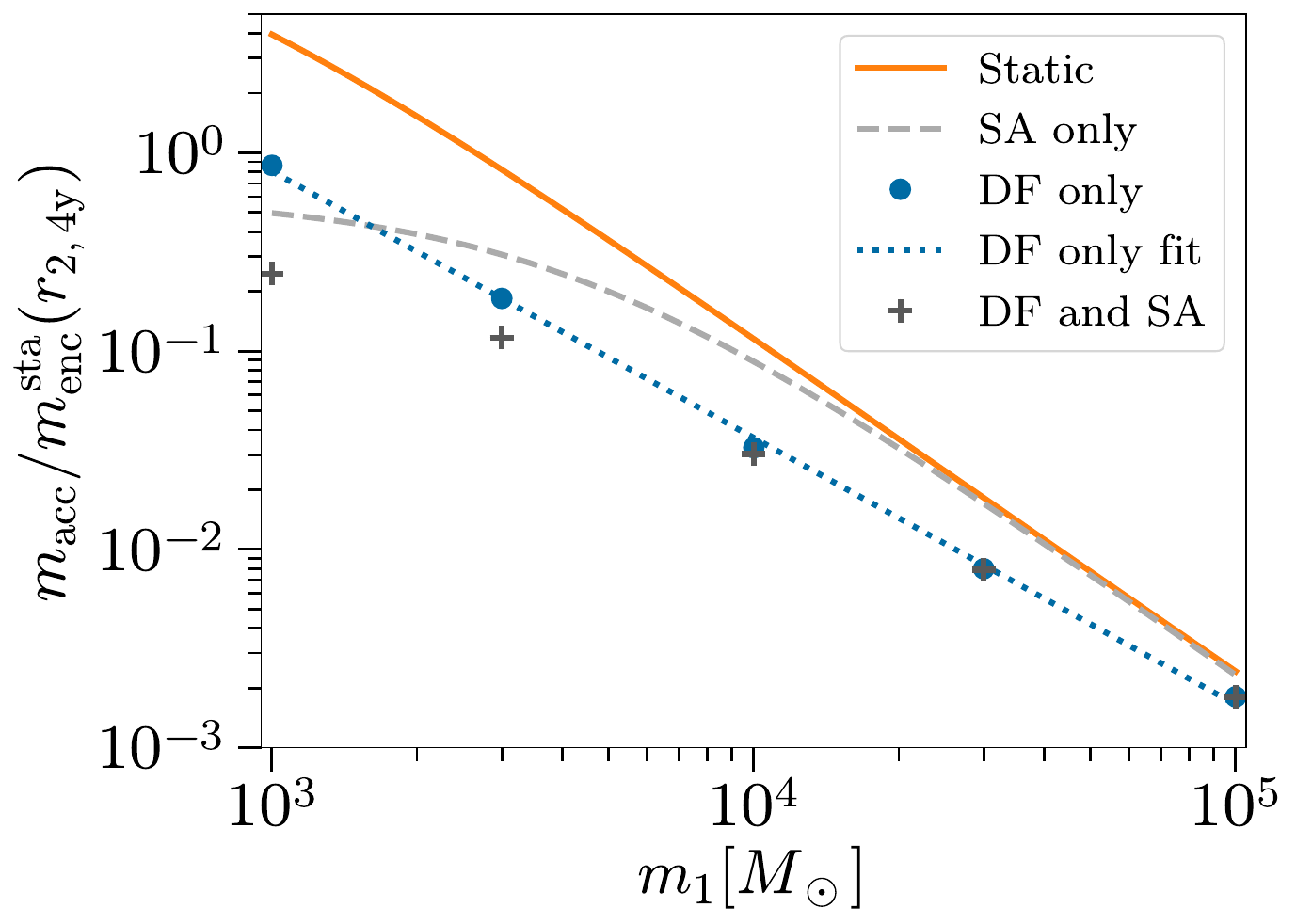}
    \caption{\textbf{The accreted mass $\mathbf{m_{acc}}$ normalized by the enclosed mass $\mathbf{m_{enc}^{sta}(r_{2,4y})}$ for different types of feedback}.
    The dark matter and binary parameters are chosen as in Fig.~\ref{fig:m_cap_by_m_enc_DF}, and the enclosed mass is that of the initial power-law distribution, also as in Fig.~\ref{fig:m_cap_by_m_enc_DF}.
    The solid orange curve is the same as the $\zeta=1$ case in Fig.~\ref{fig:m_cap_by_m_enc} with gravitational radiation reaction and dynamical friction.
    The blue circles and dotted line are the same as the dynamical-friction feedback case in Fig.~\ref{fig:m_cap_by_m_enc_DF}.
    The dashed light-gray curve corresponds to including feedback from secondary accretion only, and the dark-gray plus symbols are results of numerical simulations that include both dynamical-friction and accretion feedback.
    Further discussion of the implications of this figure are given in the text of Sec.~\ref{subsec:AllGWmacc}.}
    \label{fig:m_acc_by_m_enc_DF_SA}
\end{figure}

Next, in Fig.~\ref{fig:m_acc_by_m_enc_DF_SA}, we show the accreted mass normalized by the enclosed (static) mass of the initial dark-matter distribution for four different cases.
First, the dark gray plus signs are the visualization of the fourth column of Table~\ref{tab:DFplusSAfeed}.
Second, the blue circles and dotted line are the same results as in Fig.~\ref{fig:m_cap_by_m_enc_DF}, which corresponds to accretion onto the secondary with dynamical-friction feedback, but without feedback from secondary accretion.
Third, the dashed light-gray curve is the case in which secondary accretion and its feedback are taken into account, but dynamical friction is neglected (as in Sec.~\ref{subsec:SAonly}).
Specifically, the accreted mass was computed from the density $\rho_\DM(r,t)$, which was obtained by numerically integrating the analytical expression for the distribution function in Eq.~\eqref{eq:f_of_r2_E} over velocities.
The accreted mass was then computed through mass conservation.
Specifically, the difference in the enclosed dark-matter masses was computed when the secondary is at a separation of $r_\fy$ and when it reaches the ISCO.
The enclosed masses were computed by numerically integrating the density $\rho_\DM(r,t)$ from the inner radius to an outer radius of $30 r_\fy$; in fact, there was less of a percent difference when the upper limit of the integral was $30 r_\fy$ or $10 r_\fy$.
Fourth, the solid orange curve is the $\zeta=1$ case with both dynamical friction and radiation reaction in Fig.~\ref{fig:m_cap_by_m_enc}.
It is provided primarily for comparison with the dashed gray curve.

As suggested in Sec.~\ref{subsec:SAonly}, for larger primary masses, simulations that include secondary-accretion feedback only (no dynamical friction) overestimate the amount of accreted mass, because dynamical-friction feedback makes part of the dark matter density inaccessible to capture.
This is illustrated in Fig.~\ref{fig:m_acc_by_m_enc_DF_SA}, which shows how the SA only curve remains above the DF only curve for larger $m_1$, and in fact approaches the static curve which is a result with no feedback effects.
Given that it agrees with these static results near $m_1 = 10^5 M_\odot$, this also suggests that feedback from secondary accretion is negligible for these large primary masses.
However, for primary masses close to $10^3 M_\odot$ (mass ratios close to $10^{-2}$), including only accretion feedback leads to an accreted mass that significantly deviates from the static case, and even is a factor of two smaller than that from including only dynamical-friction feedback.
This indicates that modeling only dynamical-friction feedback for these systems leads to an inaccurate estimate of the accreted mass.

The combined effects of dynamical-friction and secondary-accretion feedback on the accreted mass are shown by the dark-gray plus symbols in Fig.~\ref{fig:m_acc_by_m_enc_DF_SA}.
For the two larger primary masses, the results are nearly indistinguishable from those with only dynamical-friction feedback.
However, for $m_1=10^4 M_\odot$, a small difference can be seen, and for $m_1=10^3 M_\odot$, using only dynamical-friction feedback leads to an overestimate of the accreted mass by roughly a factor of four.
Including secondary-accretion feedback therefore proves important to obtaining accurate estimates of the mass accreted and the impact of the secondary's inspiral on the dark-matter density after the merger (which will be the subject of the next subsection).

\begin{figure*}[t!]
    \centering
    \includegraphics[width=.49\textwidth]{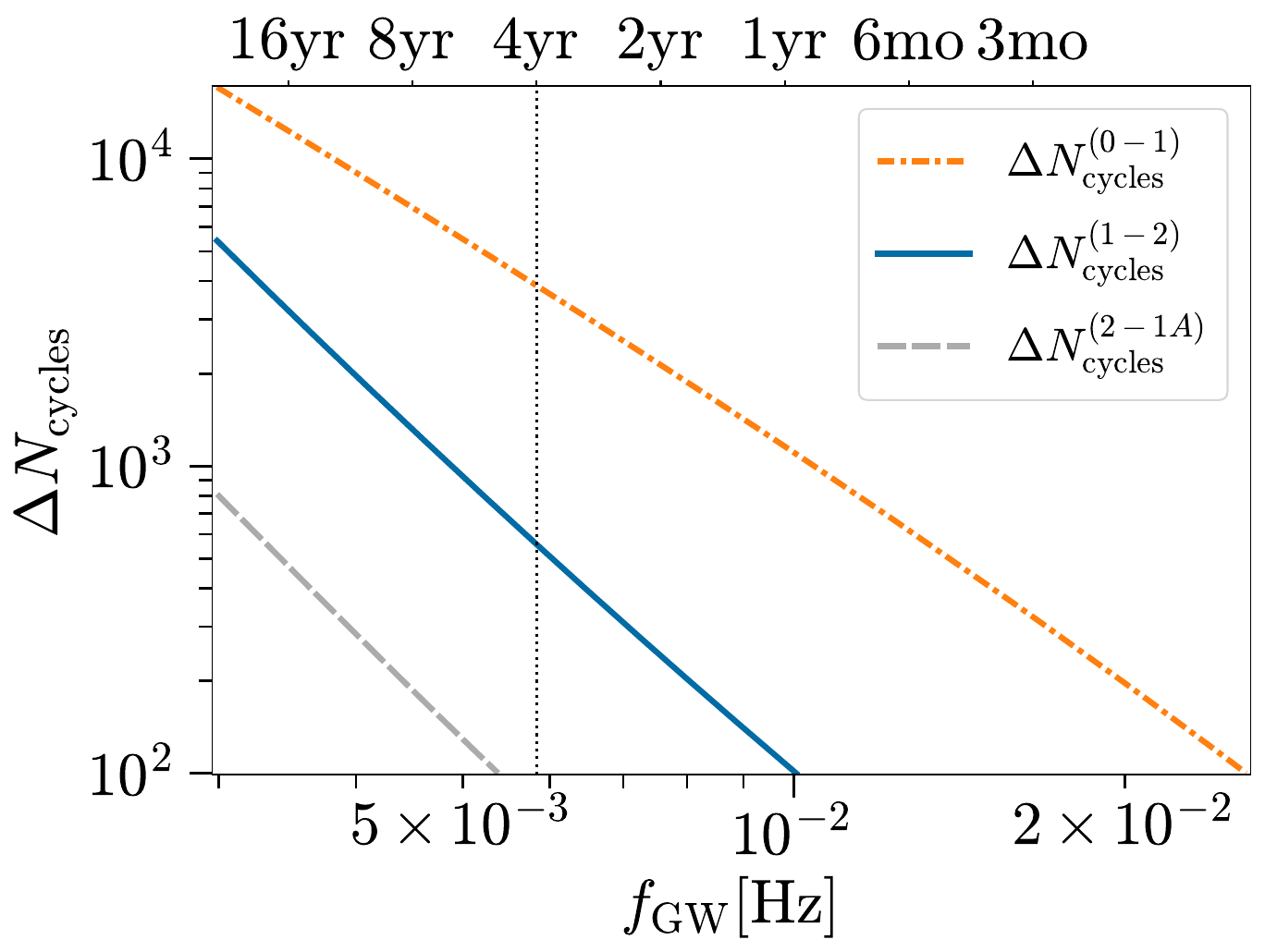} \
    \includegraphics[width=.49\textwidth]{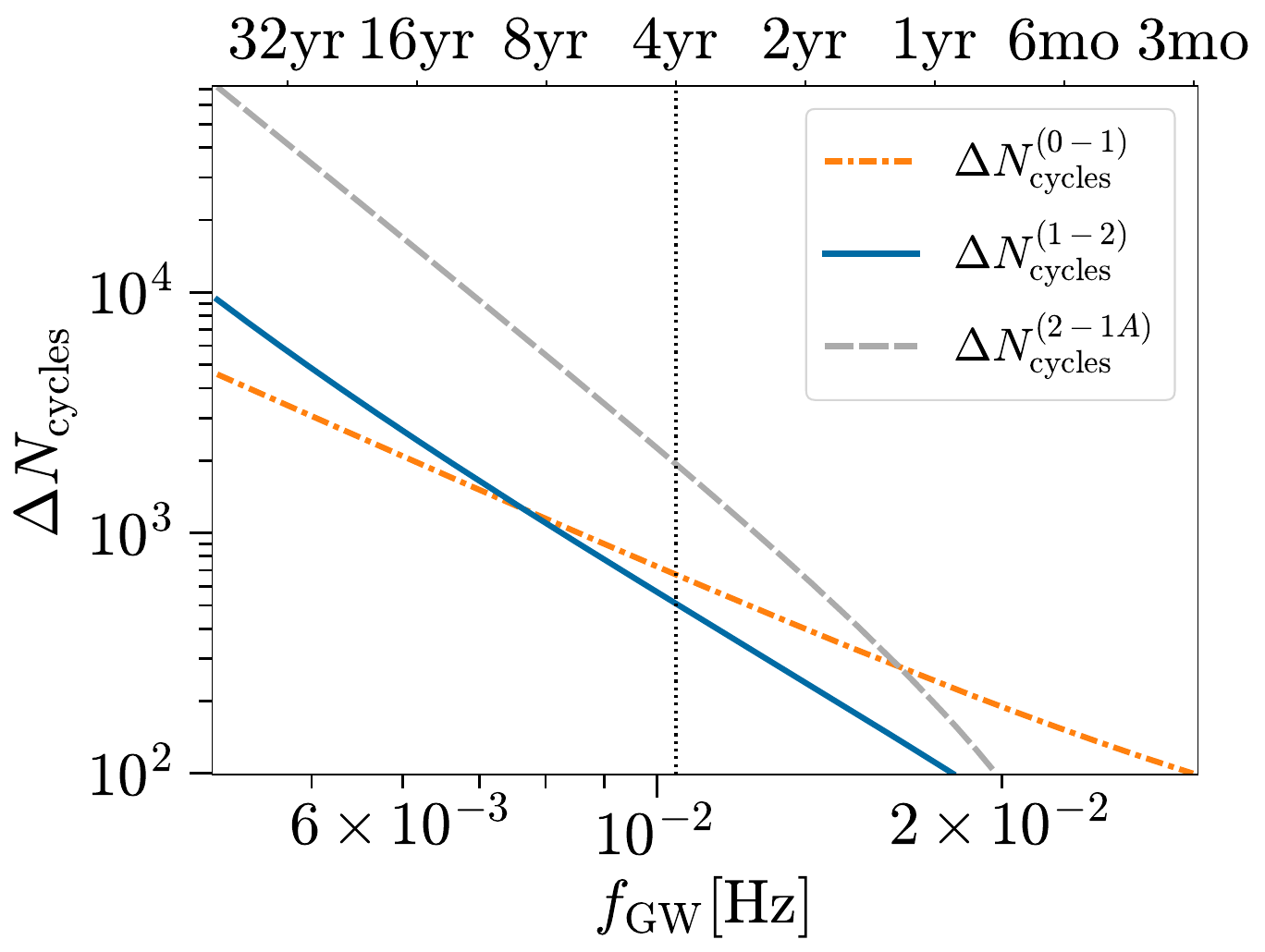}
    \caption{\textbf{Difference in the number of gravitational-wave cycles versus gravitational-wave frequency in three cases for primary masses of $\boldsymbol{10^{4} M_\odot}$ (left) and $\boldsymbol{10^{3} M_\odot}$ (right)}.
    \textit{Left}: The initial masses of the black holes are $m_1=10^4M_\odot$ and $m_2=10M_\odot$; the initial dark-matter distribution has $\rho_\SP=226M_\odot/\mathrm{pc}^3$ and $\gamma_\SP=7/3$.
    We show the number of cycles of dephasing in three different cases as a function of the initial frequency $f_{\mathrm{GW}}$, which was computed using Eqs.~\eqref{eq:Ncycles} and~\eqref{eq:DeltaNcycles}.
    The dephasing curves shown are $\Delta N^{(0-1)}_\mathrm{cycles}$ (vacuum minus dynamical-friction feedback) as the dot-dashed orange curve, $\Delta N^{(1-2)}_\mathrm{cycles}$ (dynamical-friction with feedback minus both feedback types) as the solid blue curve, and $\Delta N^{(2-1A)}_\mathrm{cycles}$ (dynamical-friction feedback and accretion without feedback minus both types of feedback)  as the dashed light-gray curve. 
    The top axis shows the time it takes for the binary to inspiral to the ISCO in vacuum from the corresponding frequency on the lower axis (with the vertical, dotted black line highlighting the four-year mark). \textit{Right}: The same as the left, except we start with a central black hole of mass $m_1=10^3M_\odot$. Further discussion of the implications of this figure are given in the text of Sec.~\ref{subsec:AllGWmacc}.}
    \label{fig:dephasing}
\end{figure*}

Figure~\ref{fig:dephasing} shows how the gravitational-wave dephasing accumulates as a function of frequency for three different dephasing comparisons and for two different primary masses: $10^4 M_\odot$ (left) and $10^3 M_\odot$ (right).
The orange dot-dashed curve, which depicts the dephasing $\Delta N^{(0-1)}_\mathrm{cycles}$ between simulations with dynamical-friction feedback and those in vacuum is similar to the curves in~\cite{Kavanagh:2020cfn}, though a larger secondary mass $m_2=10M_\odot$ is used than that in~\cite{Kavanagh:2020cfn} for the corresponding primary masses $m_1$.
The solid blue curves then show the dephasing $\Delta N^{(1-2)}_\mathrm{cycles}$ between simulations with both kinds of feedback and those with only dynamical-friction feedback.
Although $\Delta N^{(1-2)}_\mathrm{cycles}$ is smaller than $\Delta N^{(0-1)}_\mathrm{cycles}$ for all frequencies shown in the left panel of Fig.~\ref{fig:dephasing} (and comparable or smaller for separations smaller than $r_\fy$ in the right panel), the slope as a function of frequency is steeper than the slope of $\Delta N^{(0-1)}_\mathrm{cycles}$.
This suggests that the effect of accretion with feedback behaves like a more negative post-Newtonian-order effect than dynamical friction with feedback.

This last statement is worth commenting on in more detail, since for static halos, the opposite holds (the dephasing induced by accretion is a less negative post-Newtonian effect than that induced by dynamical friction, which makes the accretion dephasing less steep as a function of frequency than that of dynamical friction).
The key difference with feedback is that the different densities that contribute to the evolution of $r_2$ from dynamical friction (the local density of particles moving slower than the orbital speed) and from accretion (the local density without a restriction on speeds) have different dependencies on radius.
Because dynamical-friction feedback is more efficient, the local, effective density of more slowly moving particles becomes a steeper function of radius than both the initial density and the local density of all particles (see~\cite{Coogan:2021uqv}).
Having the density be a steeper function of radius makes the PN order of the effect less negative.
Even though dynamical friction feedback only depletes the local density of more slowly moving dark matter particles, the depletion from accretion feedback (which depletes the local density of all particles, regardless of speed) is still smaller.
This then makes this effective density less steep, and the dephasing induced by accretion a more negative PN-order effect.

This also argues for describing dynamical friction or accretion not simply in terms of powers of $r_2$ or frequency $f$, but in terms of a PN order (the number of factors of $1/c^2$) and a second small parameter, the enclosed mass ratio as a function of radius, as discussed in Sec.~\ref{subsec:IMRIeqs}.
In this classification, the PN orders of these effects are fixed, but the radial dependence of the enclosed mass ratio of all particles or of only the more slowly moving particles changes between the static and dynamic cases.
This makes it more apparent what is producing the difference in the frequency dependence of the dephasing in these cases.

The dashed light-gray curve shows the amount of dephasing with dynamical friction with feedback and  accretion without feedback from evolutions with both types of feedback (i.e., $\Delta N^{(2-1A)}_\mathrm{cycles}$).
For the $10^4 M_\odot$ primary (left), this dephasing is a factor of a few below the dephasing $\Delta N^{(1-2)}_\mathrm{cycles}$, which suggests that not including feedback overestimates the amount of dephasing, but not too significantly.
However, for the $10^3 M_\odot$ primary (right), the dephasing $\Delta N^{(2-1A)}_\mathrm{cycles}$ is significantly larger than $\Delta N^{(1-2)}_\mathrm{cycles}$, which suggests that feedback is playing an important role.
These gray curves are consistent with the numbers in Table~\ref{tab:DFplusSAfeed}, but they give a more detailed picture of how the dephasing accumulates with frequency.

\subsection{Dark-matter density} \label{subsec:AllDMdensity}

\begin{figure*}[tbh]
    \centering
    \includegraphics[width=.49\textwidth]{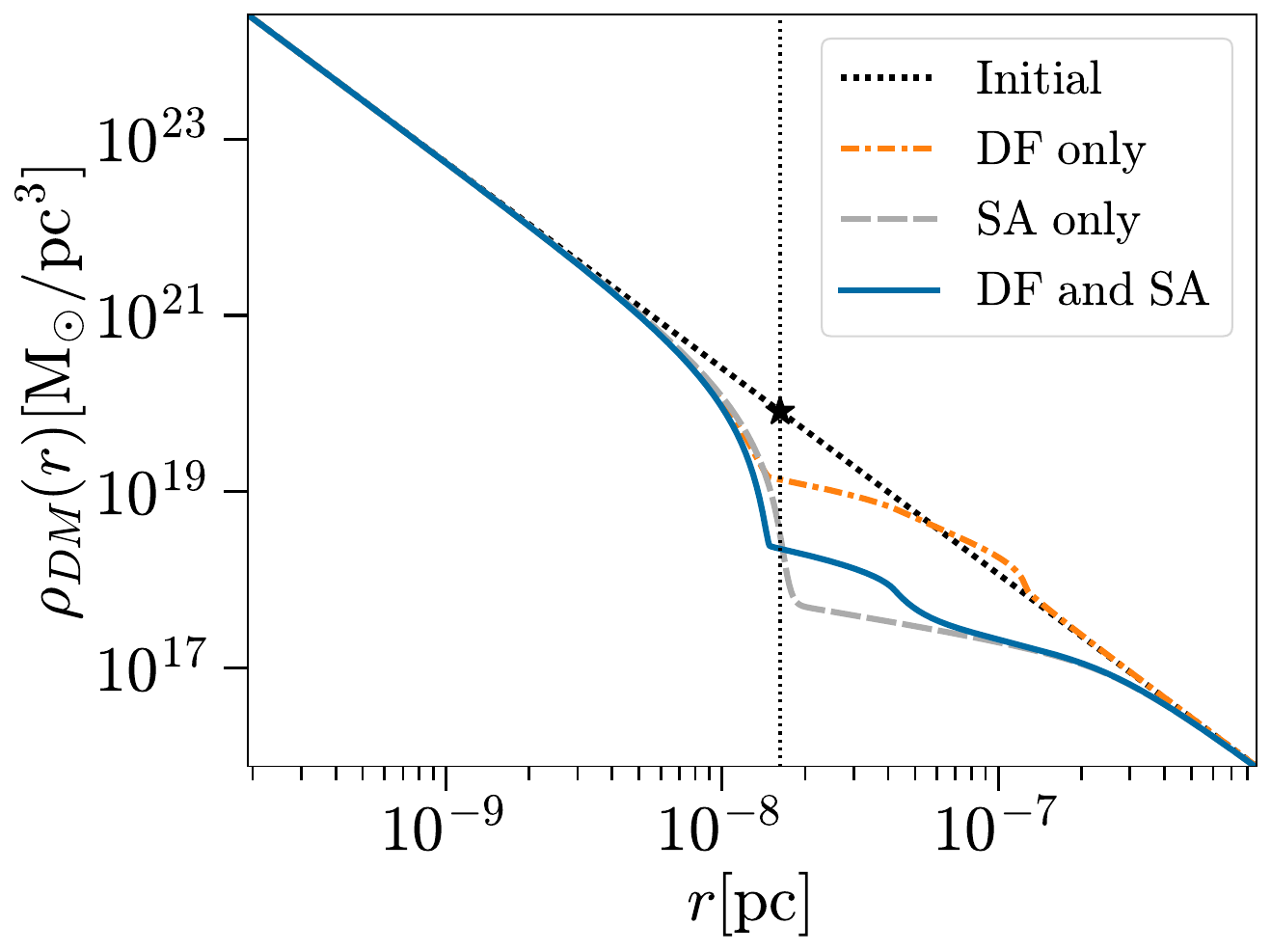} \
    \includegraphics[width=.49\textwidth]{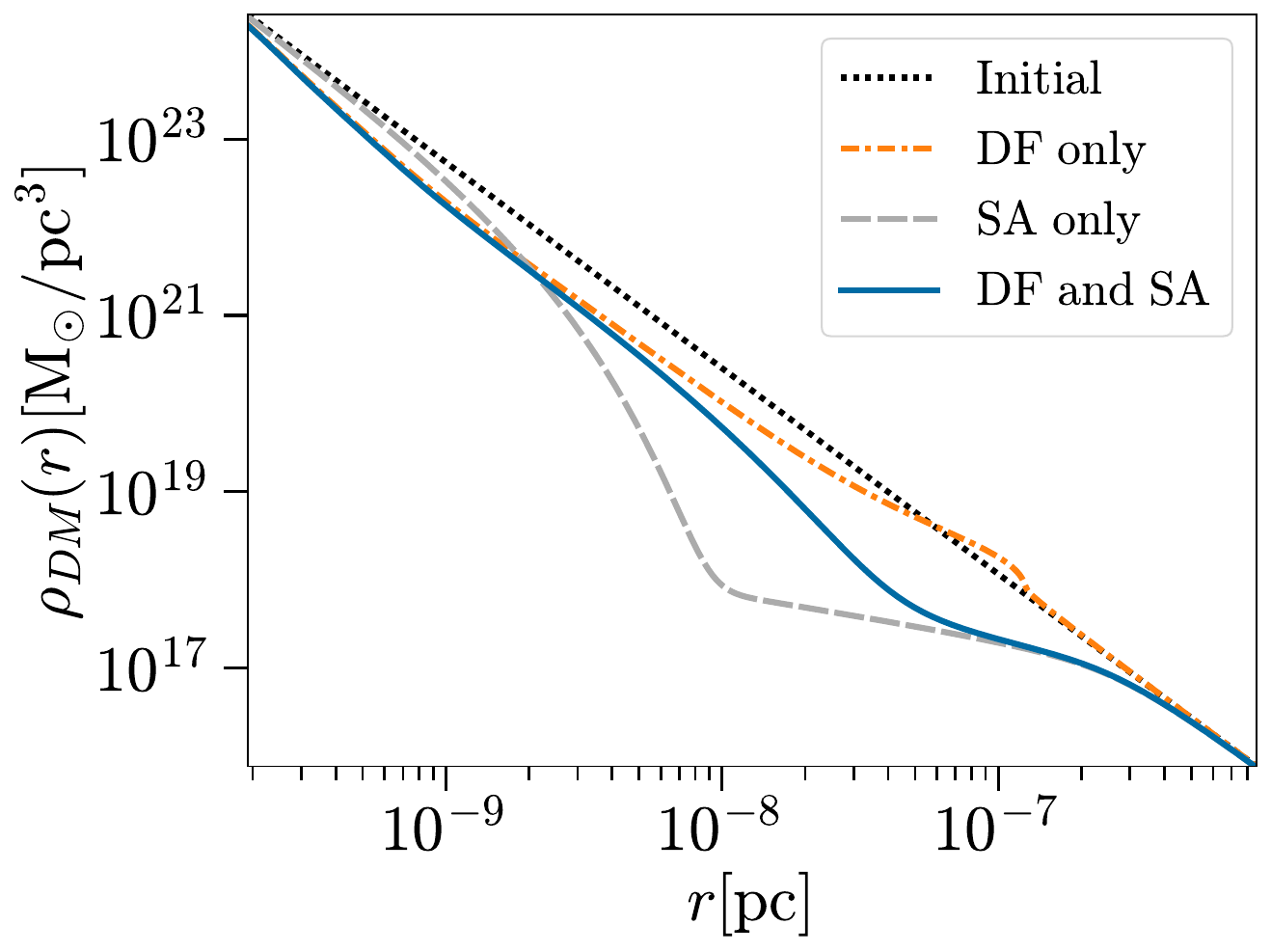}
    \caption{\textbf{Dark-matter density during (left) and after (right) the inspiral for several different cases of feedback}.
    \textit{Left}: The binary started with two black holes with masses $m_1=10^3M_\odot$ and $m_2=10M_\odot$.
    The black dotted curve is the initial dark matter distribution in Eq.~\eqref{eq:static_DM} for $\rho_\SP=226M_\odot/\mathrm{pc}^3$ and $\gamma_\SP=7/3$.
    The three other curves of different colors and line styles (dot-dashed orange, dashed light gray, and solid blue) show the densities after the binary inspirals from $3 r_\fy$ to $r_\fy$, under three different scenarios.
    The three colored curves (orange, light gray, and blue) correspond to having only dynamical friction with feedback, only secondary accretion with feedback, and both types of feedback, respectively.
    In all three cases, the binary evolution includes gravitational radiation reaction.
    In addition, for the case labeled ``DF only'' dynamical friction is also included, for ``SA only'' accretion is included but not dynamical friction, and for ``DF and SA'' all effects are included.
    The radius $r_\fy$ is shown in the left plot with a vertical black dotted line, that passes through the black star.
    \textit{Right}: The same as the left, except the density is shown after the secondary reaches the ISCO.
    Further discussion of the implications of this figure are given in the text of Sec.~\ref{subsec:AllDMdensity}.}
    \label{fig:1K_density}
\end{figure*}

In this subsection, we show the combined impact of dynamical-friction and secondary-accretion feedback on the dark-matter distribution.
The results for dynamical-friction feedback only were previously illustrated in~\cite{Kavanagh:2020cfn,Coogan:2021uqv}; those for secondary-accretion feedback only were shown in Fig.~\ref{fig:DM_capt_only}.
In Fig.~\ref{fig:1K_density}, we show all three cases (DF feedback only, SA feedback only, and both DF and SA feedback) to illustrate their respective impact on the dark matter density.

Figure~\ref{fig:1K_density} shows the density as a function of position for the three different cases of feedback (the solid blue, dashed light gray, and dash-dotted orange curves) as well as the initial density (the thick, black dotted curve) with $\rho_\SP=226M_\odot/\mathrm{pc}^3$ and $\gamma_\SP=7/3$, as before.
The case illustrated is a binary with primary mass $m_1=10^3 M_\odot$ and mass ratio $q=10^{-2}$; of the mass ratios that we considered, feedback has the most pronounced effect on the dark-matter distribution for this case.
The evolution of the binary begins at a separation or $3 r_\fy$ in both panels, but the density is shown at two different times in the binary's evolution.
In the left panel is the density when the secondary has reached the separation $r_\fy$ (as indicated by the thin black, vertical dotted line and black star), and in the right panel is the density when the secondary reaches the ISCO.

For both final radii, the curves with both types of feedback are qualitatively more similar to those of secondary accretion only at larger radii and of dynamical friction only at radii closer to the ISCO (though in the left panel, all three cases are much more similar to the initial density at these smaller radii).
In the regions where the solid blue curves are less similar to the other two cases, the density is at a value that falls somewhere between the (typically) larger DF-only curve and the smaller SA-only curve.
This behavior of the density can be understood qualitatively as follows.
When the secondary is at a fixed location, dynamical friction, in isolation, tends to increase the density at larger radii and decrease it elsewhere. 
However, as the secondary inspirals, regions that were depleted of dark matter become partially replenished. 
Secondary accretion, however, only depletes the dark matter density and does not replenish it.
Thus, as the secondary inspirals, dynamical-friction feedback makes some dark-matter particles inaccessible to accretion feedback, so that both feedback types together generally lead to a density between that of the two feedback types in isolation.
At larger radii, the inspiral is slower, which leads to a larger net decrease in the density, whereas at smaller radii, the more rapid inspiral makes feedback less effective (and thereby leads to smaller changes in the density).

In both panels, some caution should be taken in interpreting the density at the larger radii illustrated in these plots.
The region of nearly constant density that is present for only secondary-accretion feedback is sensitive to the initial radius (as illustrated in the right panel of Fig.~\ref{fig:DM_capt_only}).
Thus, that the combined feedback case asymptotes to the secondary-feedback-only case should happen independently of the choice of the initial radius of the secondary, but the specific value to which it asymptotes will depend on the initial radius.
While these figures then capture the properties of the density qualitatively, the quantitative features are sensitive to the choice of initial data.

It is also worth commenting that in the left panel, the fact that the density differs significantly from the initial density in all three cases indicates the importance of starting the evolution at $3 r_\fy$ so as to obtain ``reasonable'' initial conditions for the evolution from $r_\fy$ (where gravitational-wave emission would be strongest) that are more consistent with an adiabatic inspiral from larger radii.

\section{Discussion and conclusions} \label{sec:conclusions}

In this paper, we investigated the effect of accretion onto the secondary in intermediate mass-ratio inspirals taking place within dense distributions of dark matter.
We showed that previous calculations of the effects of the accretion on the orbital evolution of the binary were overestimates for some binaries, because they neglected the feedback from dynamical friction and from accretion.
Without any feedback, the amount of mass accreted onto the secondary could exceed the mass enclosed within the secondary's orbit.
When including feedback from dynamical friction, the amount of mass accreted still could be of the same order as the enclosed mass.
This suggested that it would be necessary to develop a method to evolve the dark-matter distribution in response to the mass removed from the dark-matter distribution via accretion.

We derived an approach to provide feedback to the dark-matter distribution from the mass accreted onto the secondary, and we proved that it satisfies mass conservation.
After implementing this method, we showed that systems without feedback do indeed overestimate the number of gravitational-wave cycles of dephasing from systems that include accretion onto the secondary but ignore feedback from accretion.
The amount of overestimation was largest for the least extreme mass ratios and became less significant at more extreme mass ratios.
Once feedback from accretion was included, the amount of dephasing induced by accretion (compared to dynamical friction) was smaller than that induced by dynamical friction (compared to vacuum), although they were of the same order at the least extreme mass ratios that we considered. 
In addition, for the less extreme mass ratios, the effects of accretion on the dark-matter density after the merger could become more significant than those of dynamical friction, particularly at larger distances from the binary's initial separation.

The frequency dependence of the dephasing induced by accretion differed significantly between static and dynamic dark-matter distributions.
For static densities, dynamical friction and accretion were negative post-Newtonian order effects, with dynamical friction being the more strongly negative.
For dynamic densities with feedback, the effects were still negative in their post-Newtonian order, but much less so.
In addition, accretion became the more negative of the two effects (a steeper dependence on frequency than dynamical friction).
This relative reversal of roles could be understood from the different properties of the enclosed mass as a function of radius of the more slowly moving particles (for dynamical friction) versus all particles (for accretion).

There are several clear directions in which the method outlined here could be extended or applied, some of which relate to the modeling of the IMRI's dynamics, others to the modeling of the gravitational waves emitted by these systems, and yet others pertaining to how well LISA could measure the effects of accretion in the emitted gravitational waves.

We begin with the aspects of the binary's dynamics.
First, the work here specialized to circular orbits, for simplicity.
However, the formation mechanisms of these IMRIs with dark matter have not been explored systematically, and the formation process is important in determining if the IMRIs form with residual eccentricity or if they would circularize once they reach the orbital separations for which LISA has the best chance of detecting their emitted gravitational waves.
Given that EMRIs without dark matter are often expected to be on eccentric orbits, it seems natural to generalize the IMRI's evolution equations to incorporate nonzero eccentricity.
Second, we used the leading Newtonian-order effects to describe the equations of motion of the IMRI.
It would also be important to formulate a relativistic description of the binary dynamics and the dark matter, so as to have a more accurate description of the binary's motion.
Third, we worked to only leading order in the mass ratio.
For IMRIs, particularly with less extreme mass ratios, it could be important to include higher-order terms in the mass-ratio expansion.
Fourth, the assumption of spherical symmetry was made in the evolution of the dark matter; this should also be revisited.
Finally, we assumed the black holes were nonspinning, but it would be of interest to consider spinning black holes, as well.
Each of these five topics would require significant new calculations, and are beyond the scope of this current work.

These improvements in computing the orbital dynamics of the binary would then make it possible to obtain more accurate predictions of the emitted gravitational waves.
In the Newtonian limit, computing the waves can be done using the quadrupole approximation to gravitational-wave emission, and the results could be obtained straightforwardly from those presented in this paper (both in the time domain or frequency domain).
Obtaining more accurate, fully relativistic waveforms, however, would be more nontrivial, as radiation reaction influences the binary's orbital dynamics; thus, the waveform generation and binary evolution equations are coupled and should be solved simultaneously.

Developing accurate gravitational-wave predictions is a necessary prerequisite for determining how well LISA could measure the presence of dark matter in IMRIs and, in addition, the effects of accretion.
The former has been investigated in~\cite{Coogan:2021uqv} using Newtonian waveforms, but the latter has yet to be studied.
An important development that allowed the detection and measurement prospects of dark matter in~\cite{Coogan:2021uqv} were approximate frequency-domain waveform models that allowed the waveform to be evaluated rapidly enough to do signal-to-noise calculations over a wider range of the dark-matter parameter space and to do parameter estimation.
Similar waveform modeling would need to be performed to do the equivalent calculations with accretion included.
Again, it would be more natural to start with Newtonian-order calculations before generalizing to fully relativistic ones.

\acknowledgments

D.A.N.\ and B.A.W.\ acknowledge support from the NSF Grants No.\ PHY-2011784 and No.\ PHY-2309021.
A.M.G.\ acknowledges the support of the Royal Society under grant number RF\textbackslash ERE\textbackslash 221005.
The authors thank Scott Hughes for pointing out Ref.~\cite{Hughes:2018qxz}.
They also acknowledge Research Computing at The University of Virginia for providing computational resources and technical support that have contributed to the results reported within this publication.


\begin{thebibliography}{32}%
\makeatletter
\providecommand \@ifxundefined [1]{%
 \@ifx{#1\undefined}
}%
\providecommand \@ifnum [1]{%
 \ifnum #1\expandafter \@firstoftwo
 \else \expandafter \@secondoftwo
 \fi
}%
\providecommand \@ifx [1]{%
 \ifx #1\expandafter \@firstoftwo
 \else \expandafter \@secondoftwo
 \fi
}%
\providecommand \natexlab [1]{#1}%
\providecommand \enquote  [1]{``#1''}%
\providecommand \bibnamefont  [1]{#1}%
\providecommand \bibfnamefont [1]{#1}%
\providecommand \citenamefont [1]{#1}%
\providecommand \href@noop [0]{\@secondoftwo}%
\providecommand \href [0]{\begingroup \@sanitize@url \@href}%
\providecommand \@href[1]{\@@startlink{#1}\@@href}%
\providecommand \@@href[1]{\endgroup#1\@@endlink}%
\providecommand \@sanitize@url [0]{\catcode `\\12\catcode `\$12\catcode
  `\&12\catcode `\#12\catcode `\^12\catcode `\_12\catcode `\%12\relax}%
\providecommand \@@startlink[1]{}%
\providecommand \@@endlink[0]{}%
\providecommand \url  [0]{\begingroup\@sanitize@url \@url }%
\providecommand \@url [1]{\endgroup\@href {#1}{\urlprefix }}%
\providecommand \urlprefix  [0]{URL }%
\providecommand \Eprint [0]{\href }%
\providecommand \doibase [0]{https://doi.org/}%
\providecommand \selectlanguage [0]{\@gobble}%
\providecommand \bibinfo  [0]{\@secondoftwo}%
\providecommand \bibfield  [0]{\@secondoftwo}%
\providecommand \translation [1]{[#1]}%
\providecommand \BibitemOpen [0]{}%
\providecommand \bibitemStop [0]{}%
\providecommand \bibitemNoStop [0]{.\EOS\space}%
\providecommand \EOS [0]{\spacefactor3000\relax}%
\providecommand \BibitemShut  [1]{\csname bibitem#1\endcsname}%
\let\auto@bib@innerbib\@empty
\bibitem [{\citenamefont {Bertone}\ and\ \citenamefont
  {Hooper}(2018)}]{Bertone:2016nfn}%
  \BibitemOpen
  \bibfield  {author} {\bibinfo {author} {\bibfnamefont {G.}~\bibnamefont
  {Bertone}}\ and\ \bibinfo {author} {\bibfnamefont {D.}~\bibnamefont
  {Hooper}},\ }\bibfield  {title} {\bibinfo {title} {{History of dark
  matter}},\ }\href {https://doi.org/10.1103/RevModPhys.90.045002} {\bibfield
  {journal} {\bibinfo  {journal} {Rev. Mod. Phys.}\ }\textbf {\bibinfo {volume}
  {90}},\ \bibinfo {pages} {045002} (\bibinfo {year} {2018})},\ \Eprint
  {https://arxiv.org/abs/1605.04909} {arXiv:1605.04909 [astro-ph.CO]}
  \BibitemShut {NoStop}%
\bibitem [{\citenamefont {Bertone}\ and\ \citenamefont
  {Tait}(2018)}]{Bertone:2018krk}%
  \BibitemOpen
  \bibfield  {author} {\bibinfo {author} {\bibfnamefont {G.}~\bibnamefont
  {Bertone}}\ and\ \bibinfo {author} {\bibfnamefont {T.}~\bibnamefont {Tait},
  \bibfnamefont {M.~P.}},\ }\bibfield  {title} {\bibinfo {title} {{A new era in
  the search for dark matter}},\ }\href
  {https://doi.org/10.1038/s41586-018-0542-z} {\bibfield  {journal} {\bibinfo
  {journal} {Nature}\ }\textbf {\bibinfo {volume} {562}},\ \bibinfo {pages}
  {51} (\bibinfo {year} {2018})},\ \Eprint {https://arxiv.org/abs/1810.01668}
  {arXiv:1810.01668 [astro-ph.CO]} \BibitemShut {NoStop}%
\bibitem [{\citenamefont {Abbott}\ \emph {et~al.}(2021)\citenamefont {Abbott}
  \emph {et~al.}}]{LIGOScientific:2021djp}%
  \BibitemOpen
  \bibfield  {author} {\bibinfo {author} {\bibfnamefont {R.}~\bibnamefont
  {Abbott}} \emph {et~al.} (\bibinfo {collaboration} {LIGO Scientific, VIRGO,
  KAGRA}),\ }\bibfield  {title} {\bibinfo {title} {{GWTC-3: Compact Binary
  Coalescences Observed by LIGO and Virgo During the Second Part of the Third
  Observing Run}},\ }\href@noop {} {\  (\bibinfo {year} {2021})},\ \Eprint
  {https://arxiv.org/abs/2111.03606} {arXiv:2111.03606 [gr-qc]} \BibitemShut
  {NoStop}%
\bibitem [{\citenamefont {Agazie}\ \emph {et~al.}(2023)\citenamefont {Agazie}
  \emph {et~al.}}]{NANOGrav:2023gor}%
  \BibitemOpen
  \bibfield  {author} {\bibinfo {author} {\bibfnamefont {G.}~\bibnamefont
  {Agazie}} \emph {et~al.} (\bibinfo {collaboration} {NANOGrav}),\ }\bibfield
  {title} {\bibinfo {title} {{The NANOGrav 15 yr Data Set: Evidence for a
  Gravitational-wave Background}},\ }\href
  {https://doi.org/10.3847/2041-8213/acdac6} {\bibfield  {journal} {\bibinfo
  {journal} {Astrophys. J. Lett.}\ }\textbf {\bibinfo {volume} {951}},\
  \bibinfo {pages} {L8} (\bibinfo {year} {2023})},\ \Eprint
  {https://arxiv.org/abs/2306.16213} {arXiv:2306.16213 [astro-ph.HE]}
  \BibitemShut {NoStop}%
\bibitem [{\citenamefont {Reardon}\ \emph {et~al.}(2023)\citenamefont {Reardon}
  \emph {et~al.}}]{Reardon:2023gzh}%
  \BibitemOpen
  \bibfield  {author} {\bibinfo {author} {\bibfnamefont {D.~J.}\ \bibnamefont
  {Reardon}} \emph {et~al.},\ }\bibfield  {title} {\bibinfo {title} {{Search
  for an Isotropic Gravitational-wave Background with the Parkes Pulsar Timing
  Array}},\ }\href {https://doi.org/10.3847/2041-8213/acdd02} {\bibfield
  {journal} {\bibinfo  {journal} {Astrophys. J. Lett.}\ }\textbf {\bibinfo
  {volume} {951}},\ \bibinfo {pages} {L6} (\bibinfo {year} {2023})},\ \Eprint
  {https://arxiv.org/abs/2306.16215} {arXiv:2306.16215 [astro-ph.HE]}
  \BibitemShut {NoStop}%
\bibitem [{\citenamefont {Antoniadis}\ \emph {et~al.}(2023)\citenamefont
  {Antoniadis} \emph {et~al.}}]{Antoniadis:2023rey}%
  \BibitemOpen
  \bibfield  {author} {\bibinfo {author} {\bibfnamefont {J.}~\bibnamefont
  {Antoniadis}} \emph {et~al.},\ }\bibfield  {title} {\bibinfo {title} {{The
  second data release from the European Pulsar Timing Array III. Search for
  gravitational wave signals}},\ }\href@noop {} {\  (\bibinfo {year} {2023})},\
  \Eprint {https://arxiv.org/abs/2306.16214} {arXiv:2306.16214 [astro-ph.HE]}
  \BibitemShut {NoStop}%
\bibitem [{\citenamefont {Bertone}\ \emph {et~al.}(2020)\citenamefont {Bertone}
  \emph {et~al.}}]{Bertone:2019irm}%
  \BibitemOpen
  \bibfield  {author} {\bibinfo {author} {\bibfnamefont {G.}~\bibnamefont
  {Bertone}} \emph {et~al.},\ }\bibfield  {title} {\bibinfo {title}
  {{Gravitational wave probes of dark matter: challenges and opportunities}},\
  }\href {https://doi.org/10.21468/SciPostPhysCore.3.2.007} {\bibfield
  {journal} {\bibinfo  {journal} {SciPost Phys. Core}\ }\textbf {\bibinfo
  {volume} {3}},\ \bibinfo {pages} {007} (\bibinfo {year} {2020})},\ \Eprint
  {https://arxiv.org/abs/1907.10610} {arXiv:1907.10610 [astro-ph.CO]}
  \BibitemShut {NoStop}%
\bibitem [{\citenamefont {Baker}\ \emph {et~al.}(2019)\citenamefont {Baker}
  \emph {et~al.}}]{Baker:2019nia}%
  \BibitemOpen
  \bibfield  {author} {\bibinfo {author} {\bibfnamefont {J.}~\bibnamefont
  {Baker}} \emph {et~al.},\ }\bibfield  {title} {\bibinfo {title} {{The Laser
  Interferometer Space Antenna: Unveiling the Millihertz Gravitational Wave
  Sky}},\ }\href@noop {} {\  (\bibinfo {year} {2019})},\ \Eprint
  {https://arxiv.org/abs/1907.06482} {arXiv:1907.06482 [astro-ph.IM]}
  \BibitemShut {NoStop}%
\bibitem [{\citenamefont {Eda}\ \emph {et~al.}(2013)\citenamefont {Eda},
  \citenamefont {Itoh}, \citenamefont {Kuroyanagi},\ and\ \citenamefont
  {Silk}}]{Eda:2013gg}%
  \BibitemOpen
  \bibfield  {author} {\bibinfo {author} {\bibfnamefont {K.}~\bibnamefont
  {Eda}}, \bibinfo {author} {\bibfnamefont {Y.}~\bibnamefont {Itoh}}, \bibinfo
  {author} {\bibfnamefont {S.}~\bibnamefont {Kuroyanagi}},\ and\ \bibinfo
  {author} {\bibfnamefont {J.}~\bibnamefont {Silk}},\ }\bibfield  {title}
  {\bibinfo {title} {{New Probe of Dark-Matter Properties: Gravitational Waves
  from an Intermediate-Mass Black Hole Embedded in a Dark-Matter Minispike}},\
  }\href {https://doi.org/10.1103/PhysRevLett.110.221101} {\bibfield  {journal}
  {\bibinfo  {journal} {Phys. Rev. Lett.}\ }\textbf {\bibinfo {volume} {110}},\
  \bibinfo {pages} {221101} (\bibinfo {year} {2013})},\ \Eprint
  {https://arxiv.org/abs/1301.5971} {arXiv:1301.5971 [gr-qc]} \BibitemShut
  {NoStop}%
\bibitem [{\citenamefont {Eda}\ \emph {et~al.}(2015)\citenamefont {Eda},
  \citenamefont {Itoh}, \citenamefont {Kuroyanagi},\ and\ \citenamefont
  {Silk}}]{Eda:2014kra}%
  \BibitemOpen
  \bibfield  {author} {\bibinfo {author} {\bibfnamefont {K.}~\bibnamefont
  {Eda}}, \bibinfo {author} {\bibfnamefont {Y.}~\bibnamefont {Itoh}}, \bibinfo
  {author} {\bibfnamefont {S.}~\bibnamefont {Kuroyanagi}},\ and\ \bibinfo
  {author} {\bibfnamefont {J.}~\bibnamefont {Silk}},\ }\bibfield  {title}
  {\bibinfo {title} {{Gravitational waves as a probe of dark matter
  minispikes}},\ }\href {https://doi.org/10.1103/PhysRevD.91.044045} {\bibfield
   {journal} {\bibinfo  {journal} {Phys. Rev. D}\ }\textbf {\bibinfo {volume}
  {91}},\ \bibinfo {pages} {044045} (\bibinfo {year} {2015})},\ \Eprint
  {https://arxiv.org/abs/1408.3534} {arXiv:1408.3534 [gr-qc]} \BibitemShut
  {NoStop}%
\bibitem [{\citenamefont {Yue}\ and\ \citenamefont {Han}(2018)}]{Yue:2017iwc}%
  \BibitemOpen
  \bibfield  {author} {\bibinfo {author} {\bibfnamefont {X.-J.}\ \bibnamefont
  {Yue}}\ and\ \bibinfo {author} {\bibfnamefont {W.-B.}\ \bibnamefont {Han}},\
  }\bibfield  {title} {\bibinfo {title} {{Gravitational waves with dark matter
  minispikes: the combined effect}},\ }\href
  {https://doi.org/10.1103/PhysRevD.97.064003} {\bibfield  {journal} {\bibinfo
  {journal} {Phys. Rev. D}\ }\textbf {\bibinfo {volume} {97}},\ \bibinfo
  {pages} {064003} (\bibinfo {year} {2018})},\ \Eprint
  {https://arxiv.org/abs/1711.09706} {arXiv:1711.09706 [gr-qc]} \BibitemShut
  {NoStop}%
\bibitem [{\citenamefont {Yue}\ \emph {et~al.}(2019)\citenamefont {Yue},
  \citenamefont {Han},\ and\ \citenamefont {Chen}}]{Yue:2018vtk}%
  \BibitemOpen
  \bibfield  {author} {\bibinfo {author} {\bibfnamefont {X.-J.}\ \bibnamefont
  {Yue}}, \bibinfo {author} {\bibfnamefont {W.-B.}\ \bibnamefont {Han}},\ and\
  \bibinfo {author} {\bibfnamefont {X.}~\bibnamefont {Chen}},\ }\bibfield
  {title} {\bibinfo {title} {{Dark matter: an efficient catalyst for
  intermediate-mass-ratio-inspiral events}},\ }\href
  {https://doi.org/10.3847/1538-4357/ab06f6} {\bibfield  {journal} {\bibinfo
  {journal} {Astrophys. J.}\ }\textbf {\bibinfo {volume} {874}},\ \bibinfo
  {pages} {34} (\bibinfo {year} {2019})},\ \Eprint
  {https://arxiv.org/abs/1802.03739} {arXiv:1802.03739 [gr-qc]} \BibitemShut
  {NoStop}%
\bibitem [{\citenamefont {Edwards}\ \emph {et~al.}(2020)\citenamefont
  {Edwards}, \citenamefont {Chianese}, \citenamefont {Kavanagh}, \citenamefont
  {Nissanke},\ and\ \citenamefont {Weniger}}]{Edwards:2019tzf}%
  \BibitemOpen
  \bibfield  {author} {\bibinfo {author} {\bibfnamefont {T.~D.~P.}\
  \bibnamefont {Edwards}}, \bibinfo {author} {\bibfnamefont {M.}~\bibnamefont
  {Chianese}}, \bibinfo {author} {\bibfnamefont {B.~J.}\ \bibnamefont
  {Kavanagh}}, \bibinfo {author} {\bibfnamefont {S.~M.}\ \bibnamefont
  {Nissanke}},\ and\ \bibinfo {author} {\bibfnamefont {C.}~\bibnamefont
  {Weniger}},\ }\bibfield  {title} {\bibinfo {title} {{Unique Multimessenger
  Signal of QCD Axion Dark Matter}},\ }\href
  {https://doi.org/10.1103/PhysRevLett.124.161101} {\bibfield  {journal}
  {\bibinfo  {journal} {Phys. Rev. Lett.}\ }\textbf {\bibinfo {volume} {124}},\
  \bibinfo {pages} {161101} (\bibinfo {year} {2020})},\ \Eprint
  {https://arxiv.org/abs/1905.04686} {arXiv:1905.04686 [hep-ph]} \BibitemShut
  {NoStop}%
\bibitem [{\citenamefont {Kavanagh}\ \emph {et~al.}(2020)\citenamefont
  {Kavanagh}, \citenamefont {Nichols}, \citenamefont {Bertone},\ and\
  \citenamefont {Gaggero}}]{Kavanagh:2020cfn}%
  \BibitemOpen
  \bibfield  {author} {\bibinfo {author} {\bibfnamefont {B.~J.}\ \bibnamefont
  {Kavanagh}}, \bibinfo {author} {\bibfnamefont {D.~A.}\ \bibnamefont
  {Nichols}}, \bibinfo {author} {\bibfnamefont {G.}~\bibnamefont {Bertone}},\
  and\ \bibinfo {author} {\bibfnamefont {D.}~\bibnamefont {Gaggero}},\
  }\bibfield  {title} {\bibinfo {title} {{Detecting dark matter around black
  holes with gravitational waves: Effects of dark-matter dynamics on the
  gravitational waveform}},\ }\href
  {https://doi.org/10.1103/PhysRevD.102.083006} {\bibfield  {journal} {\bibinfo
   {journal} {Phys. Rev. D}\ }\textbf {\bibinfo {volume} {102}},\ \bibinfo
  {pages} {083006} (\bibinfo {year} {2020})},\ \Eprint
  {https://arxiv.org/abs/2002.12811} {arXiv:2002.12811 [gr-qc]} \BibitemShut
  {NoStop}%
\bibitem [{\citenamefont {Coogan}\ \emph {et~al.}(2022)\citenamefont {Coogan},
  \citenamefont {Bertone}, \citenamefont {Gaggero}, \citenamefont {Kavanagh},\
  and\ \citenamefont {Nichols}}]{Coogan:2021uqv}%
  \BibitemOpen
  \bibfield  {author} {\bibinfo {author} {\bibfnamefont {A.}~\bibnamefont
  {Coogan}}, \bibinfo {author} {\bibfnamefont {G.}~\bibnamefont {Bertone}},
  \bibinfo {author} {\bibfnamefont {D.}~\bibnamefont {Gaggero}}, \bibinfo
  {author} {\bibfnamefont {B.~J.}\ \bibnamefont {Kavanagh}},\ and\ \bibinfo
  {author} {\bibfnamefont {D.~A.}\ \bibnamefont {Nichols}},\ }\bibfield
  {title} {\bibinfo {title} {{Measuring the dark matter environments of black
  hole binaries with gravitational waves}},\ }\href
  {https://doi.org/10.1103/PhysRevD.105.043009} {\bibfield  {journal} {\bibinfo
   {journal} {Phys. Rev. D}\ }\textbf {\bibinfo {volume} {105}},\ \bibinfo
  {pages} {043009} (\bibinfo {year} {2022})},\ \Eprint
  {https://arxiv.org/abs/2108.04154} {arXiv:2108.04154 [gr-qc]} \BibitemShut
  {NoStop}%
\bibitem [{\citenamefont {Becker}\ \emph {et~al.}(2022)\citenamefont {Becker},
  \citenamefont {Sagunski}, \citenamefont {Prinz},\ and\ \citenamefont
  {Rastgoo}}]{Becker:2021ivq}%
  \BibitemOpen
  \bibfield  {author} {\bibinfo {author} {\bibfnamefont {N.}~\bibnamefont
  {Becker}}, \bibinfo {author} {\bibfnamefont {L.}~\bibnamefont {Sagunski}},
  \bibinfo {author} {\bibfnamefont {L.}~\bibnamefont {Prinz}},\ and\ \bibinfo
  {author} {\bibfnamefont {S.}~\bibnamefont {Rastgoo}},\ }\bibfield  {title}
  {\bibinfo {title} {{Circularization versus eccentrification in intermediate
  mass ratio inspirals inside dark matter spikes}},\ }\href
  {https://doi.org/10.1103/PhysRevD.105.063029} {\bibfield  {journal} {\bibinfo
   {journal} {Phys. Rev. D}\ }\textbf {\bibinfo {volume} {105}},\ \bibinfo
  {pages} {063029} (\bibinfo {year} {2022})},\ \Eprint
  {https://arxiv.org/abs/2112.09586} {arXiv:2112.09586 [gr-qc]} \BibitemShut
  {NoStop}%
\bibitem [{\citenamefont {Becker}\ and\ \citenamefont
  {Sagunski}(2022)}]{Becker:2022wlo}%
  \BibitemOpen
  \bibfield  {author} {\bibinfo {author} {\bibfnamefont {N.}~\bibnamefont
  {Becker}}\ and\ \bibinfo {author} {\bibfnamefont {L.}~\bibnamefont
  {Sagunski}},\ }\bibfield  {title} {\bibinfo {title} {{Comparing Accretion
  Disks and Dark Matter Spikes in Intermediate Mass Ratio Inspirals}},\
  }\href@noop {} {\  (\bibinfo {year} {2022})},\ \Eprint
  {https://arxiv.org/abs/2211.05145} {arXiv:2211.05145 [gr-qc]} \BibitemShut
  {NoStop}%
\bibitem [{\citenamefont {Speeney}\ \emph {et~al.}(2022)\citenamefont
  {Speeney}, \citenamefont {Antonelli}, \citenamefont {Baibhav},\ and\
  \citenamefont {Berti}}]{Speeney:2022ryg}%
  \BibitemOpen
  \bibfield  {author} {\bibinfo {author} {\bibfnamefont {N.}~\bibnamefont
  {Speeney}}, \bibinfo {author} {\bibfnamefont {A.}~\bibnamefont {Antonelli}},
  \bibinfo {author} {\bibfnamefont {V.}~\bibnamefont {Baibhav}},\ and\ \bibinfo
  {author} {\bibfnamefont {E.}~\bibnamefont {Berti}},\ }\bibfield  {title}
  {\bibinfo {title} {{Impact of relativistic corrections on the detectability
  of dark-matter spikes with gravitational waves}},\ }\href
  {https://doi.org/10.1103/PhysRevD.106.044027} {\bibfield  {journal} {\bibinfo
   {journal} {Phys. Rev. D}\ }\textbf {\bibinfo {volume} {106}},\ \bibinfo
  {pages} {044027} (\bibinfo {year} {2022})},\ \Eprint
  {https://arxiv.org/abs/2204.12508} {arXiv:2204.12508 [gr-qc]} \BibitemShut
  {NoStop}%
\bibitem [{\citenamefont {Cole}\ \emph
  {et~al.}(2022{\natexlab{a}})\citenamefont {Cole}, \citenamefont {Bertone},
  \citenamefont {Coogan}, \citenamefont {Gaggero}, \citenamefont {Karydas},
  \citenamefont {Kavanagh}, \citenamefont {Spieksma},\ and\ \citenamefont
  {Tomaselli}}]{Cole:2022fir}%
  \BibitemOpen
  \bibfield  {author} {\bibinfo {author} {\bibfnamefont {P.~S.}\ \bibnamefont
  {Cole}}, \bibinfo {author} {\bibfnamefont {G.}~\bibnamefont {Bertone}},
  \bibinfo {author} {\bibfnamefont {A.}~\bibnamefont {Coogan}}, \bibinfo
  {author} {\bibfnamefont {D.}~\bibnamefont {Gaggero}}, \bibinfo {author}
  {\bibfnamefont {T.}~\bibnamefont {Karydas}}, \bibinfo {author} {\bibfnamefont
  {B.~J.}\ \bibnamefont {Kavanagh}}, \bibinfo {author} {\bibfnamefont
  {T.~F.~M.}\ \bibnamefont {Spieksma}},\ and\ \bibinfo {author} {\bibfnamefont
  {G.~M.}\ \bibnamefont {Tomaselli}},\ }\bibfield  {title} {\bibinfo {title}
  {{Disks, spikes, and clouds: distinguishing environmental effects on BBH
  gravitational waveforms}},\ }\href@noop {} {\  (\bibinfo {year}
  {2022}{\natexlab{a}})},\ \Eprint {https://arxiv.org/abs/2211.01362}
  {arXiv:2211.01362 [gr-qc]} \BibitemShut {NoStop}%
\bibitem [{\citenamefont {Cole}\ \emph
  {et~al.}(2022{\natexlab{b}})\citenamefont {Cole}, \citenamefont {Coogan},
  \citenamefont {Kavanagh},\ and\ \citenamefont {Bertone}}]{Cole:2022ucw}%
  \BibitemOpen
  \bibfield  {author} {\bibinfo {author} {\bibfnamefont {P.~S.}\ \bibnamefont
  {Cole}}, \bibinfo {author} {\bibfnamefont {A.}~\bibnamefont {Coogan}},
  \bibinfo {author} {\bibfnamefont {B.~J.}\ \bibnamefont {Kavanagh}},\ and\
  \bibinfo {author} {\bibfnamefont {G.}~\bibnamefont {Bertone}},\ }\bibfield
  {title} {\bibinfo {title} {{Measuring dark matter spikes around primordial
  black holes with Einstein Telescope and Cosmic Explorer}},\ }\href@noop {} {\
   (\bibinfo {year} {2022}{\natexlab{b}})},\ \Eprint
  {https://arxiv.org/abs/2207.07576} {arXiv:2207.07576 [astro-ph.CO]}
  \BibitemShut {NoStop}%
\bibitem [{\citenamefont {Barausse}\ \emph {et~al.}(2014)\citenamefont
  {Barausse}, \citenamefont {Cardoso},\ and\ \citenamefont
  {Pani}}]{Barausse:2014tra}%
  \BibitemOpen
  \bibfield  {author} {\bibinfo {author} {\bibfnamefont {E.}~\bibnamefont
  {Barausse}}, \bibinfo {author} {\bibfnamefont {V.}~\bibnamefont {Cardoso}},\
  and\ \bibinfo {author} {\bibfnamefont {P.}~\bibnamefont {Pani}},\ }\bibfield
  {title} {\bibinfo {title} {{Can environmental effects spoil precision
  gravitational-wave astrophysics?}},\ }\href
  {https://doi.org/10.1103/PhysRevD.89.104059} {\bibfield  {journal} {\bibinfo
  {journal} {Phys. Rev. D}\ }\textbf {\bibinfo {volume} {89}},\ \bibinfo
  {pages} {104059} (\bibinfo {year} {2014})},\ \Eprint
  {https://arxiv.org/abs/1404.7149} {arXiv:1404.7149 [gr-qc]} \BibitemShut
  {NoStop}%
\bibitem [{\citenamefont {Macedo}\ \emph {et~al.}(2013)\citenamefont {Macedo},
  \citenamefont {Pani}, \citenamefont {Cardoso},\ and\ \citenamefont
  {Crispino}}]{Macedo:2013qea}%
  \BibitemOpen
  \bibfield  {author} {\bibinfo {author} {\bibfnamefont {C.~F.~B.}\
  \bibnamefont {Macedo}}, \bibinfo {author} {\bibfnamefont {P.}~\bibnamefont
  {Pani}}, \bibinfo {author} {\bibfnamefont {V.}~\bibnamefont {Cardoso}},\ and\
  \bibinfo {author} {\bibfnamefont {L.~C.~B.}\ \bibnamefont {Crispino}},\
  }\bibfield  {title} {\bibinfo {title} {{Into the lair: gravitational-wave
  signatures of dark matter}},\ }\href
  {https://doi.org/10.1088/0004-637X/774/1/48} {\bibfield  {journal} {\bibinfo
  {journal} {Astrophys. J.}\ }\textbf {\bibinfo {volume} {774}},\ \bibinfo
  {pages} {48} (\bibinfo {year} {2013})},\ \Eprint
  {https://arxiv.org/abs/1302.2646} {arXiv:1302.2646 [gr-qc]} \BibitemShut
  {NoStop}%
\bibitem [{\citenamefont {Chandrasekhar}(1943)}]{Chandrasekhar1943a}%
  \BibitemOpen
  \bibfield  {author} {\bibinfo {author} {\bibfnamefont {S.}~\bibnamefont
  {Chandrasekhar}},\ }\bibfield  {title} {\bibinfo {title} {{Dynamical
  Friction. I. General Considerations: the Coefficient of Dynamical
  Friction.}},\ }\href {https://doi.org/10.1086/144517} {\bibfield  {journal}
  {\bibinfo  {journal} {The Astrophysical Journal}\ }\textbf {\bibinfo {volume}
  {97}},\ \bibinfo {pages} {255} (\bibinfo {year} {1943})}\BibitemShut
  {NoStop}%
\bibitem [{\citenamefont {Gondolo}\ and\ \citenamefont
  {Silk}(1999)}]{Gondolo:1999ef}%
  \BibitemOpen
  \bibfield  {author} {\bibinfo {author} {\bibfnamefont {P.}~\bibnamefont
  {Gondolo}}\ and\ \bibinfo {author} {\bibfnamefont {J.}~\bibnamefont {Silk}},\
  }\bibfield  {title} {\bibinfo {title} {{Dark matter annihilation at the
  galactic center}},\ }\href {https://doi.org/10.1103/PhysRevLett.83.1719}
  {\bibfield  {journal} {\bibinfo  {journal} {Phys. Rev. Lett.}\ }\textbf
  {\bibinfo {volume} {83}},\ \bibinfo {pages} {1719} (\bibinfo {year}
  {1999})},\ \Eprint {https://arxiv.org/abs/astro-ph/9906391}
  {arXiv:astro-ph/9906391} \BibitemShut {NoStop}%
\bibitem [{\citenamefont {Ullio}\ \emph {et~al.}(2001)\citenamefont {Ullio},
  \citenamefont {Zhao},\ and\ \citenamefont {Kamionkowski}}]{Ullio:2001fb}%
  \BibitemOpen
  \bibfield  {author} {\bibinfo {author} {\bibfnamefont {P.}~\bibnamefont
  {Ullio}}, \bibinfo {author} {\bibfnamefont {H.}~\bibnamefont {Zhao}},\ and\
  \bibinfo {author} {\bibfnamefont {M.}~\bibnamefont {Kamionkowski}},\
  }\bibfield  {title} {\bibinfo {title} {{A Dark matter spike at the galactic
  center?}},\ }\href {https://doi.org/10.1103/PhysRevD.64.043504} {\bibfield
  {journal} {\bibinfo  {journal} {Phys. Rev. D}\ }\textbf {\bibinfo {volume}
  {64}},\ \bibinfo {pages} {043504} (\bibinfo {year} {2001})},\ \Eprint
  {https://arxiv.org/abs/astro-ph/0101481} {arXiv:astro-ph/0101481}
  \BibitemShut {NoStop}%
\bibitem [{\citenamefont {Sadeghian}\ \emph {et~al.}(2013)\citenamefont
  {Sadeghian}, \citenamefont {Ferrer},\ and\ \citenamefont
  {Will}}]{Sadeghian:2013laa}%
  \BibitemOpen
  \bibfield  {author} {\bibinfo {author} {\bibfnamefont {L.}~\bibnamefont
  {Sadeghian}}, \bibinfo {author} {\bibfnamefont {F.}~\bibnamefont {Ferrer}},\
  and\ \bibinfo {author} {\bibfnamefont {C.~M.}\ \bibnamefont {Will}},\
  }\bibfield  {title} {\bibinfo {title} {{Dark matter distributions around
  massive black holes: A general relativistic analysis}},\ }\href
  {https://doi.org/10.1103/PhysRevD.88.063522} {\bibfield  {journal} {\bibinfo
  {journal} {Phys. Rev. D}\ }\textbf {\bibinfo {volume} {88}},\ \bibinfo
  {pages} {063522} (\bibinfo {year} {2013})},\ \Eprint
  {https://arxiv.org/abs/1305.2619} {arXiv:1305.2619 [astro-ph.GA]}
  \BibitemShut {NoStop}%
\bibitem [{\citenamefont {Barack}\ and\ \citenamefont
  {Pound}(2019)}]{Barack:2018yvs}%
  \BibitemOpen
  \bibfield  {author} {\bibinfo {author} {\bibfnamefont {L.}~\bibnamefont
  {Barack}}\ and\ \bibinfo {author} {\bibfnamefont {A.}~\bibnamefont {Pound}},\
  }\bibfield  {title} {\bibinfo {title} {{Self-force and radiation reaction in
  general relativity}},\ }\href {https://doi.org/10.1088/1361-6633/aae552}
  {\bibfield  {journal} {\bibinfo  {journal} {Rept. Prog. Phys.}\ }\textbf
  {\bibinfo {volume} {82}},\ \bibinfo {pages} {016904} (\bibinfo {year}
  {2019})},\ \Eprint {https://arxiv.org/abs/1805.10385} {arXiv:1805.10385
  [gr-qc]} \BibitemShut {NoStop}%
\bibitem [{\citenamefont {Peters}\ and\ \citenamefont
  {Mathews}(1963)}]{Peters:1963ux}%
  \BibitemOpen
  \bibfield  {author} {\bibinfo {author} {\bibfnamefont {P.~C.}\ \bibnamefont
  {Peters}}\ and\ \bibinfo {author} {\bibfnamefont {J.}~\bibnamefont
  {Mathews}},\ }\bibfield  {title} {\bibinfo {title} {{Gravitational radiation
  from point masses in a Keplerian orbit}},\ }\href
  {https://doi.org/10.1103/PhysRev.131.435} {\bibfield  {journal} {\bibinfo
  {journal} {Phys. Rev.}\ }\textbf {\bibinfo {volume} {131}},\ \bibinfo {pages}
  {435} (\bibinfo {year} {1963})}\BibitemShut {NoStop}%
\bibitem [{\citenamefont {Peters}(1964)}]{Peters:1964zz}%
  \BibitemOpen
  \bibfield  {author} {\bibinfo {author} {\bibfnamefont {P.~C.}\ \bibnamefont
  {Peters}},\ }\bibfield  {title} {\bibinfo {title} {{Gravitational Radiation
  and the Motion of Two Point Masses}},\ }\href
  {https://doi.org/10.1103/PhysRev.136.B1224} {\bibfield  {journal} {\bibinfo
  {journal} {Phys. Rev.}\ }\textbf {\bibinfo {volume} {136}},\ \bibinfo {pages}
  {B1224} (\bibinfo {year} {1964})}\BibitemShut {NoStop}%
\bibitem [{\citenamefont {Hughes}(2019)}]{Hughes:2018qxz}%
  \BibitemOpen
  \bibfield  {author} {\bibinfo {author} {\bibfnamefont {S.~A.}\ \bibnamefont
  {Hughes}},\ }\bibfield  {title} {\bibinfo {title} {{Bound orbits of a slowly
  evolving black hole}},\ }\href {https://doi.org/10.1103/PhysRevD.100.064001}
  {\bibfield  {journal} {\bibinfo  {journal} {Phys. Rev. D}\ }\textbf {\bibinfo
  {volume} {100}},\ \bibinfo {pages} {064001} (\bibinfo {year} {2019})},\
  \Eprint {https://arxiv.org/abs/1806.09022} {arXiv:1806.09022 [gr-qc]}
  \BibitemShut {NoStop}%
\bibitem [{\citenamefont {Unruh}(1976)}]{Unruh:1976fm}%
  \BibitemOpen
  \bibfield  {author} {\bibinfo {author} {\bibfnamefont {W.~G.}\ \bibnamefont
  {Unruh}},\ }\bibfield  {title} {\bibinfo {title} {{Absorption Cross-Section
  of Small Black Holes}},\ }\href {https://doi.org/10.1103/PhysRevD.14.3251}
  {\bibfield  {journal} {\bibinfo  {journal} {Phys. Rev. D}\ }\textbf {\bibinfo
  {volume} {14}},\ \bibinfo {pages} {3251} (\bibinfo {year}
  {1976})}\BibitemShut {NoStop}%
\bibitem [{Hal(2022)}]{HaloFeedback}%
  \BibitemOpen
  \href@noop {} {}\bibinfo {howpublished}
  {\url{https://github.com/bradkav/HaloFeedback}} (\bibinfo {year} {2022}),\
  \bibinfo {note} {\textsc{HaloFeedback} [code]}\BibitemShut {NoStop}%
\end{thebibliography}
\end{document}